\documentclass[prb,twocolumn,nofootinbib,superscriptaddress,aps]{revtex4-2}
\usepackage{savesym}
\usepackage[T1]{fontenc}
\let\hbar\relax
\usepackage{times}
\usepackage{xcolor}
\usepackage{stmaryrd}
\usepackage{bbold}
\usepackage{currvita}
\usepackage{graphicx}
\usepackage{dcolumn}
\usepackage{bm}
\usepackage{slashed}
\usepackage{tikz} 
\usetikzlibrary{decorations.pathmorphing}
\usetikzlibrary{decorations.markings}
\usetikzlibrary{arrows.meta}
\usepackage{extramarks}
\usepackage{amsmath}
\usepackage[caption=false]{subfig}
\usepackage{amsthm}
\usepackage{amsfonts}
\usepackage{mathrsfs}
\usepackage{color}
\usepackage{booktabs}
\usepackage{mathtools}
\usepackage{tensor, amsmath, amssymb}
\usepackage{mathdots}
\usepackage[symbol*]{footmisc}
\newcommand{\state}[1]{|#1\rangle}

\newcommand{\ms}{\mathsf}

\newcommand{\tocless}[2]{\bgroup\let\addcontentsline=\nocontentsline#1{#2}\egroup}
\newcommand{\Chi}{\mathrm{X}}

\usepackage[breaklinks=true
,colorlinks
,citecolor=bred
,linkcolor=bleu
,urlcolor=bleu]{hyperref}
\definecolor{bleu}{rgb}{0.0, 0.5, 0.69}
\definecolor{bred}{rgb}{0.8, 0.25, 0.33}

\tikzset{snake it/.style={decorate, decoration=snake}}
{\def\subfigLabel{#1}\begin{tabular}[b]{@{}c@{}}}%
{\\[10pt]\subfigLabel\end{tabular}}
\usepackage{tikz}
\usepackage{tablefootnote}
\usetikzlibrary{decorations.markings}

\newsavebox{\TempBox}

\theoremstyle{definition} 
\begin{document}
\renewcommand*{\thefootnote}{\fnsymbol{footnote}}

\title{Anomalous gapped boundaries between surface topological orders \\in higher-order topological insulators and superconductors with inversion symmetry}

\author{Ming-Hao Li}

\affiliation{Rudolf Peierls Centre for Theoretical Physics, Clarendon Laboratory, University of Oxford, Oxford, OX1 3PU, UK}
\affiliation{Department of Physics, University of Zurich, Winterthurerstrasse 190, 8057 Zurich, Switzerland}
\author{Titus Neupert}
\affiliation{Department of Physics, University of Zurich, Winterthurerstrasse 190, 8057 Zurich, Switzerland}
\author{S. A. Parameswaran}
\affiliation{Rudolf Peierls Centre for Theoretical Physics, Clarendon Laboratory, University of Oxford, Oxford, OX1 3PU, UK}
\author{Apoorv Tiwari}
\affiliation{Department of Physics, University of Zurich, Winterthurerstrasse 190, 8057 Zurich, Switzerland}
\affiliation{Condensed Matter Theory Group, Paul Scherrer Institute, CH-5232 Villigen PSI, Switzerland}
\affiliation{Department of Physics, KTH Royal Institute of Technology, Stockholm, 106 91 Sweden}

\date{\today}
\begin{abstract}
We show that the gapless boundary signatures --- namely, chiral/helical hinge modes or localized zero modes ---  of three-dimensional higher-order topological
insulators and superconductors with inversion symmetry can be gapped without symmetry breaking upon the introduction of non-Abelian surface topological order. In each  case, the fractionalization pattern that appears on the surface is `anomalous' in the sense that it can be made consistent with symmetry only on the surface of a three dimensional higher-order
insulator/superconductor. Our results show that  the interacting manifestation of higher-order topology is the appearance of `anomalous gapped boundaries' between distinct topological orders whose quasiparticles are related by inversion, possibly in conjunction with other protecting symmetries such as TRS and charge conservation.
\end{abstract}
\maketitle

\section{Introduction}\label{Sec:intro}
Over the past decades there have been great strides in the classification and characterization of topological phases of matter \cite{Kane-Mele-2005, Kane-Mele-2007, bernevig_zhang, RMP_ChingKai,Schnyder_2008, Kitaev_2009, cohomology1, DW_ashvin, cohomology2, wanggu, kapustin2015fermionic, freed2021reflection, maissam_new_5, bhardwaj2017state, cheng_bi_you_gu, maissam_new_3, Aasen:2021vva}.
Distinctions between these phases are encoded in the entanglement structure of their quantum wave functions rather than in patterns of broken symmetry \cite{Oshikawa_2010, senthil_2013, zeng2019quantum}. 
While some gapless systems such as topological semimetals \cite{{Savrasov_2011, Zyuzin_2012}} and quantum spin liquids \cite{Savary_2016, sl-review-2} exhibit topological characteristics, in this work, we focus exclusively on topological phases with a bulk energy gap, which often coexists with gapless modes localized on system boundaries.
Such gapped topological phases fall into distinct equivalence classes that cannot be adiabatically deformed into each other without encountering a phase transition at which the bulk gap closes. 
In some cases, these distinctions rely on the presence of protecting global symmetry(ies); if such symmetries are broken --- either spontaneously or explicitly --- then phases may be deformed into each other without encountering a phase boundary.
Such phases are said to be `symmetry protected topological' (SPT) phases \cite{cohomology1, cohomology2}.
Another class of phases does {\it not} rely on such symmetry protection, and are said to have `intrinsic' topological order \cite{Kitaev_toric_code, wen2004quantum, Levin_Wen}. 
A  separate distinction can be usefully drawn between invertible topological phases --- those with no non-trivial topological excitations --- and non-invertible phases which host such excitations.  
Invertible topological order can thus either be intrinsic, as in the chiral $p_x + ip_y$ superconductor \cite{Read_Green}, or symmetry-protected, as exemplified by three dimensional (3D) time-reversal invariant topological band insulators (TIs), which rely on a combination of $\ms{U}(1)$ particle number conservation  and time-reversal symmetry (TRS) $\mathcal T$ \cite{Kane-Mele-2007}. 
Non-invertible topological orders have to be intrinsic and  are then distinguished from each other and from trivial orders by the  quantum statistics or braiding properties of their topological excitations. 
However such features can be `enriched' by global symmetries, which allow finer distinctions to be made between distinct patterns of quantum number fractionalization \cite{Wang_2019}.
Examples of gapped non-invertible topological orders include Kitaev's toric code \cite{Kitaev_toric_code}, Abelian and non-Abelian fractional quantum Hall states \cite{Wen_1995, xiao-gang_zee, Read_Rezayi, Mooreread2}, and quantum spin liquids \cite{Savary_2016, sl-review-2, PUTROV2017254, Xiao_2016, Apoorv_2017}. 

Invertible and non-invertible topological orders are linked through  the notion of `anomalous' fractionalization, a focus of this paper. An invertible SPT with purely on-site symmetries in $d$ dimensions generically has gapless  modes on  $d-1$ dimensional boundaries that  respect the protecting symmetries.  These gapless modes are  anomalous in that they cannot be realized in a strictly $d-1$ dimensional system equipped  with the same symmetries, and therefore require the bulk in order  to exist --- a feature often termed the `bulk-boundary correspondence' \cite{Ryu_2012, Chang_Tse_2014, kapustin2014symmetry, Chang_Tse_2016, Han_2017, Tiwari_2018, Witten_2016, witten_path}. 
One route to gapping these modes in the absence of a bulk phase transition involves  symmetry breaking, but an additional possibility emerges  on  the two-dimensional (2D) surface of a 3D SPT:  namely, the formation of a non-invertible  2D topological order, enriched by the same symmetries \cite{Senthil_2013b}.
This  necessarily involves interactions, since non-invertible orders are intrinsically interacting. The bulk-boundary correspondence is now  encoded in the fact that the resulting symmetry-enriched topological (SET) order is {\it also} anomalous: its fractionalized quasiparticles transform under symmetry in a manner that is impossible in a strictly two-dimensional system, but is admissible on the 2D surface of a 3D SPT \cite{Senthil_2013b, Chen_2014, Bonderson_2013}. A specific example of this is furnished by the 3D TRS invariant topological insulator \cite{Kane-Mele-2007}. The gapless `surface termination'  that preserves symmetry is a single 2D Dirac fermion, which would violate theorems on `fermion doubling' were it to appear in a purely 2D TRS lattice system \cite{NIELSEN1981219, Witten_2016, witten_path}. The second, gapped, possibility is the non-invertible $\mathcal T$-Pfaffian topological order which contains non-Abelian anyonic excitations \cite{Chen_2014, Bonderson_2013}. Notably, despite respecting TRS $\mathcal T$, the anyon content of the $\mathcal T$-Pfaffian requires a non-zero chiral central charge; this is incompatible with $\mathcal T$-symmetry in a strictly 2D system, but can be realized in a $\mathcal T$-preserving manner on the 2D surface of the 3DTI. Similar gapped anomalous surface topological orders (STOs) have been proposed for many bosonic and fermionic SPT phases \cite{Chen_2015, Fidkowski_2013, barkeshli2017, Barkeshli:2016mew, Meng_2018, kobayashi2019gapped, juven_wen_witten, Tachikawa_TRS, Seiberg:2016rsg, juven_wen_witten, maissam_new_1, maissam_new_2} and have been used as the basis of a classification of interacting electronic topological insulators.

The introduction of spatial symmetries adds richness to the topological classification, allowing the identification of  crystalline \cite{fu_crystalline, isobe_fu, fu_review, titus_lecture, gauging_crystalline} and `higher order' symmetry-protected topological phases  (HOTPs) of matter~\cite{benalcazar-bernevig-hughes-prb-2017,Schindlereaat-2018,Yuxuan-2018, Piet-2017, Chen-Fang-2019, Peterson_2018, Serra_Garcia_2018, imhof2018topolectrical, luka_2019, yizhi_hoint_2018, shiozaki_2019, bultinck_2019, Ingham_2020, rasmussen_2020}. A topological crystalline insulator/superconductor, as its name implies, requires one or more crystalline symmetries, and exhibits gapless states only on surfaces that preserve those symmetries. However, generic surfaces may preserve symmetries only within certain high-symmetry subsystems, e.g., reflection symmetric lines on a 2D surface, or rotation symmetric points on a 1D or 2D edge; different patches of surfaces can  be mapped into each other under symmetry \cite{Hermele_1, Hermele_2, Hermele_3}. These facts can complicate a straightforward definition of  a signature on a $d-1$ dimensional surface as it may not fully realize all the symmetries of the bulk. Higher-order topology resolves this complication, by identifying robust  signatures on `boundaries of boundaries' and generalizations thereof. An $n^\text{th}$ order topological phase of matter in $d$ dimensions hosts gapless excitations on $d-n$ dimensional boundary subsystems. In the   $d=3$ case of interest to us, a first-order topological phase has gapless 2D surface  states, a second-order phase is gapless along 1D high-symmetry lines of its 2D surface, and a third-order phase has gapless modes localized to 0D points on its surface. (The latter  two cases are often termed `hinges' or `corners', reflecting their spatial locations when the protecting symmetry is a point-group). A large class of  HOSPTs have been identified in fermionic and bosonic systems, and candidate solid-state materials have been proposed to host gapless modes protected by higher-order topology.

In 3D, the concept of anomalous surface topological order also generalizes to HOSPTs, but in a distinct fashion from the $n=1$ case. This was demonstrated in Ref.~\onlinecite{apoorv_2019} in the specific setting of HOSPTs protected by a combination of ${\ms C}_{2n}$ rotations and ${\mathcal T}$, which host gapless chiral modes on the hinges of a ${\ms C}_{2n}$-symmetric sample. It was demonstrated that a consistent STO for ${\ms C}_{2n}{\mathcal T}$ HOSPTs could be generated by placing a cousin of the $\mathcal T$-Pfaffian topological order, with the same anyons and symmetry transformation properties, but with  $\mathcal T$-symmetry broken in  two opposite senses, on adjacent patches of the surface that get mapped into each other under the action of ${\ms C}_{2n}$. In a purely 2D setting --- imagine these phases `painted' on a hollow ${\ms C}_{2n}$-symmetric shell --- this pattern   would necessarily involve chiral boundary modes between topological orders with distinct senses of $\mathcal{T}$-breaking. Their absence ---and hence the presence of an {\it anomalous gapped boundary} --- is because the gapless modes can be gapped while preserving symmetry when combined with  those contributed by the HOSPT bulk, which counter-propagate and have the same symmetry properties. Thus, the manifestation of higher order topology in this strongly-interacting setting is through the anomalous  gapped boundaries of a certain symmetry-enforced patterning of topological orders.
[We note  there have been further subtle notions introduced which distinguish bulk-boundary phenomena in the context of {\it weakly} interacting electronic systems, such as boundary-obstructed topological phases \cite{Khalaf_Benalcazar_Hughes_Queiroz_2019, tiwari2020chiral,Ezawa_2020, Ronny_2020, Jahin_2022}, phases with obstructed atomic limits, and fragile topology \cite{Bradlyn_2017,Po_2018, Cano_2018, Po_2019, Khalaf_PRX, Bradlyn_2019, iraola2021topological, rjslager_1, rjslager_2};  we focus on strongly interacting, non-fragile phases and do not consider these below.] 

\begin{table*}[ht]\label{Table: Main}
\begin{tabular}{ c|c|c||c|c } 
\toprule
\hline
 & \multicolumn{2}{c||}{Second Order Topology} & \multicolumn{2}{c}{Third Order Topology}\\
 \hline
AZ Class & Hinge Modes& STOs&Zero Modes& STOs \\
\hline
A & Chiral Dirac & 2D $\mathcal{T}$-Pfaffian $^\dagger$ & &  \\
\hline
AIII & &  & Dirac & $(\mathsf{SO}(3)_3)^4$ $^\ddagger$
 \\
\hline
AI &  &  & &  \\
\hline
BDI &  & & Majorana & $(\mathsf{SO}(3)_3)^2 $ \\
\hline
D & Chiral Majorana & $\mathsf{SO}(3)_3 $ & Majorana & $(\mathsf{SO}(3)_3)^2 $ \\
\hline
DIII & Helical Majorana  & $\mathsf{SO}(3)_3 \times \overline{\mathsf{SO}(3)_3}$ & Majorana Kramers Pair& $(\mathsf{SO}(3)_3 \times \overline{\mathsf{SO}(3)_3})^2$  \\
\hline
AII & Helical Dirac & 2D $\mathcal{T}$-Pfaffian$\times \overline{\text{2D $\mathcal{T}$-Pfaffian}}$  & &  \\
\hline
CII &  &  & Majorana Kramers Pair & $(\mathsf{SO}(3)_3 \times \overline{\mathsf{SO}(3)_3})^2$ \\
\hline
C& Chiral Majorana &  $^\S$
& &  \\
\hline
CI &  &  & &  \\
\hline
\bottomrule
\multicolumn{5}{l}{\footnotesize $^\dagger$ The 2D $\mathcal{T}$-Pfaffian has the same anyon content of $\mathcal{T}$-Pfaffian, albeit without TRS.} \\
\multicolumn{5}{l}{\footnotesize $^\ddagger$ Also enriched by $\mathsf{U}(1)$ charge conservation symmetry.} \\

\multicolumn{5}{l}{\footnotesize $^\S$ Note that there is no STO for 2nd order class C. The 2nd order class C can be obtained by breaking the TRS in the first order class CI,  } \\ 
\multicolumn{5}{l}{\footnotesize similar to the case of 2nd order class A which is obtained by breaking the TRS in the first order class AII. The STO for the 2nd order class C}\\
\multicolumn{5}{l}{\footnotesize should have the same anyon content as the STO for the first order class CI, which was excluded in Ref.~\onlinecite{ci-noSTO}. In the above paper, the authors}\\
 \multicolumn{5}{l}{\footnotesize argue that, due to disorder, interaction is always relevant in the first order class CI,and will cause spontaneous symmetry breaking of TRS, } \\
\multicolumn{5}{l}{\footnotesize thus ruling out STOs that preserve TRS.} \\
\end{tabular}
\caption{\label{tab}Summary of surface topological order for all inversion symmetric higher order topological phases in the AZ classes. The superscripts that appear in the column of third order STO denote the number of copies of the topological order e.g. $(\mathsf{SO}(3)_3)^2$ means two copies of $\mathsf{SO}(3)_3$. The STO we put on the surface is always inversion symmetric since we place a topological order and its inversion symmetric partner on the surface so that the original gapless line/point modes can be gapped.}
\end{table*}

In this work, we construct topologically ordered surface terminations for three dimensional electronic topological insulators and superconductors both with and without TRS (classes A, AII, AIII, D, DIII, CII and BDI within the Altland-Zirnbauer classification scheme) whose higher-order topology  is enabled by the additional presence of three-dimensional spatial inversion symmetry, denoted $\mathcal{I}$. A band-theoretic classification  indicates that the surfaces of such phases host one-dimensional chiral or helical Dirac hinge modes along an inversion-invariant  line, or   degenerate zero-dimensional corner modes at antipodal points, which cannot be gapped by any symmetric free-fermion perturbation \cite{khalaf_inversion}. We show that these hinge and corner modes may be  gapped out upon introducing an inversion-symmetric configuration of fractionalized phases with non-Abelian anyons on the surface. The surface now realizes a fully gapped and symmetric topologically ordered state. Crucially, this surface fractionalization pattern is anomalous as it would be impossible to assemble  a configuration of topological orders with the relevant symmetry properties for a system in purely two dimensions, i.e. without invoking the mode contributed by the three dimensional bulk. Compared with Ref.~\onlinecite{apoorv_2019}, this work discusses the action of crystalline symmetry on the STO beyond merely arranging the STOs in a symmetry-respecting configuration.

The rest of the paper is organized as follows. In Section \ref{Sec:2ndOrder} and \ref{Sec:3rdOrder}, we present the construction of surface topological order for second and third order inversion symmetric topological phases respectively. The main results of the paper are summarised in Table \ref{Table: Main}. Technical details are collected in several appendices. 

\section{Surface Topological Order for Second Order Topological Phases} \label{Sec:2ndOrder}
\subsection{Class A $+$ Inversion: HOTI with Chiral Dirac Hinge Mode} \label{Sec:HOTI}
\begin{figure}[tbh!]
\centering
\subfloat[Before pasting STO]{%
  \includegraphics[width=.44\linewidth]{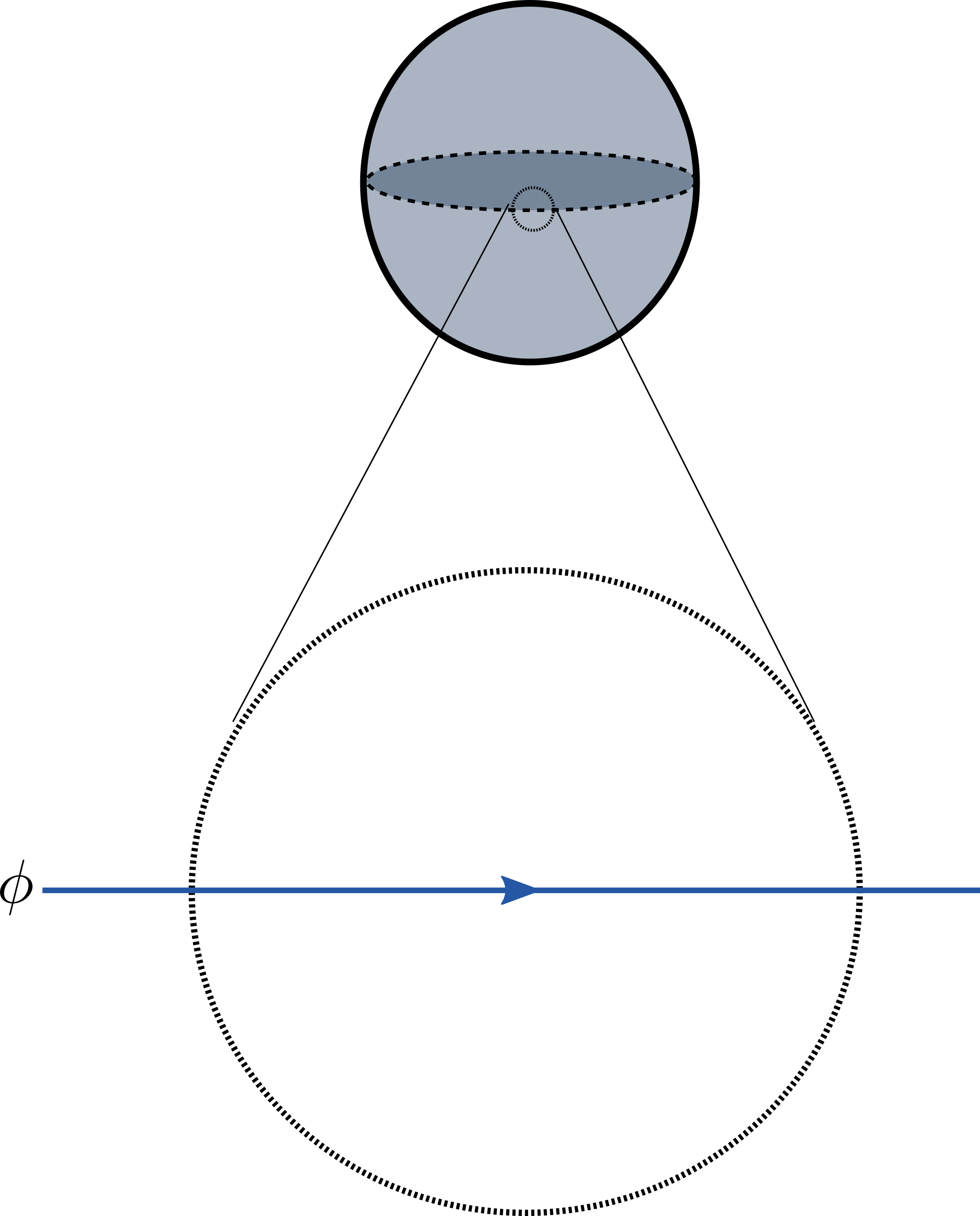}%
}\hfill
\subfloat[After pasting STO]{%
  \includegraphics[width=.49\linewidth]{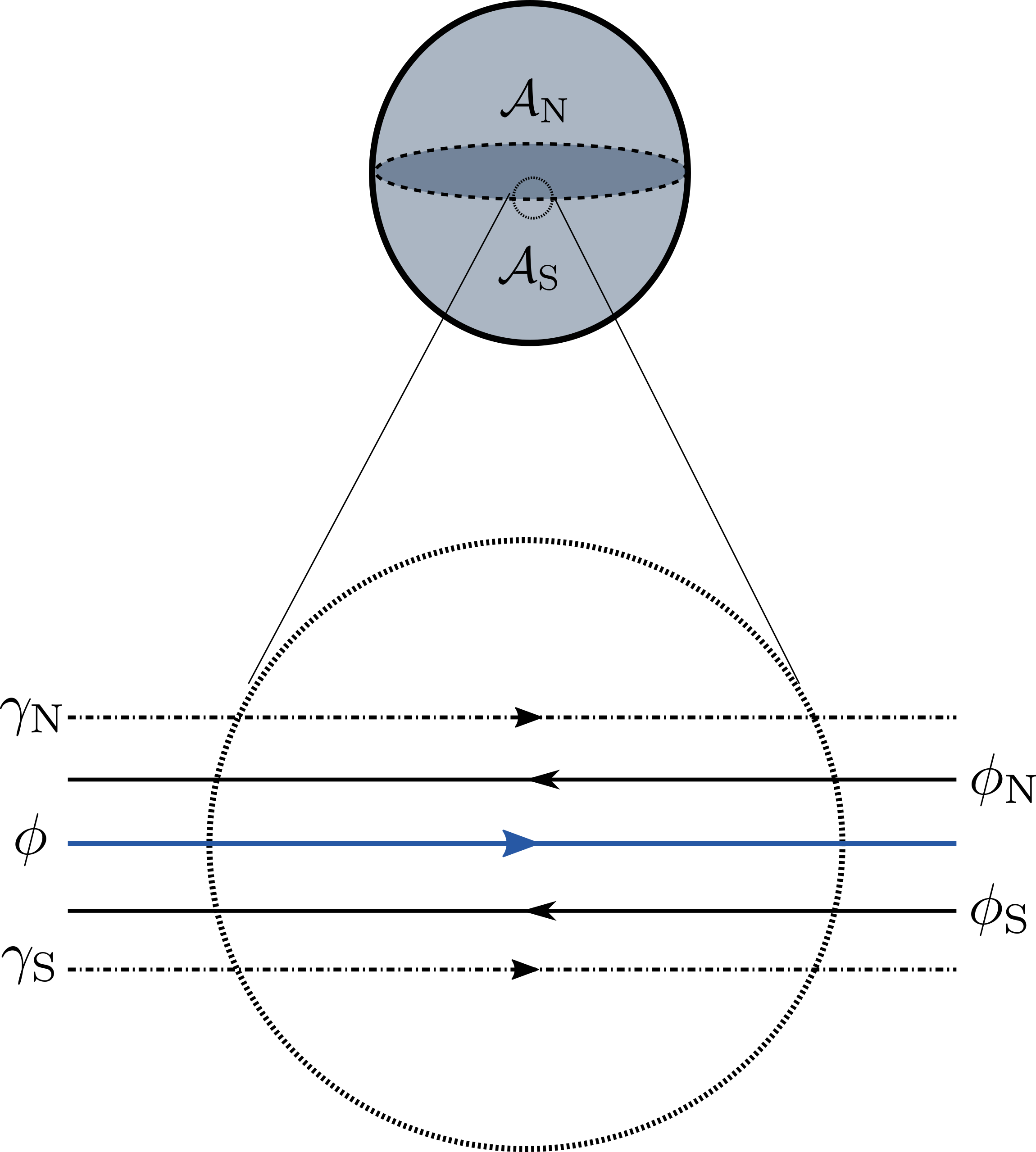}%
}
\caption{Surface topological order for second order topological phases protected by inversion symmetry in A class.}
\label{figure_for_sto}
\end{figure}
A second-order topological insulator (HOTI) protected solely by inversion symmetry can be obtained by perturbing a 3D topological insulator with a surface mass term that breaks TRS ($\mathcal{T}$) but preserves inversion symmetry ($\mathcal{I}$) \cite{khalaf_inversion}. The bulk of the 3D topological insulator can be captured by the Bloch Hamiltonian
\begin{equation}
H(\vec{k})=\sum_{i}\sin(k_i)\sigma_i\otimes \tau_x-(2-\sum_{i}\cos(k_i))\sigma_0\otimes \tau_z,
\end{equation}
where $\vec{\sigma}, \vec{\tau}$ are Pauli matrices that act on spin and orbital degrees of freedom respectively. Inversion symmetry and TRS are given by $\mathcal{I}=\sigma_0\otimes \tau_z$ and $\mathcal{T}=i\sigma_y \otimes \tau_0\mathcal{K}$ respectively, where $\mathcal{K}$ denotes complex conjugation. In order to construct the  hinge state  we first inspect the linearized surface Hamiltonian which takes the form \cite{khalaf_inversion}
\begin{equation}
h(\vec{k},\vec{r})=-(\vec{k}\times \hat{n}_{\vec{r}})\cdot \vec{\tilde{\sigma}},
\end{equation}
where $\hat{n}_{\vec{r}}$ is the unit vector normal to the surface, which is assumed to be spherical. The projected surface Hamiltonian is expressed in a rotated $\tilde{\sigma}$ basis in which the symmetry operators act as $\mathcal{I}=-\tilde{\sigma}_0$ and $\mathcal{T}=i\tilde{\sigma}_y \mathcal{K}$. The surface hinge mode is obtained by perturbing the 3D TI with a TRS breaking mass term of the form $\delta h(\vec{r})=m_{\vec{r}}(\hat{n}_{\vec{r}}\cdot \vec{\tilde{\sigma}})$.
Inversion symmetry imposes the constraint $m_{\vec{r}}=-m_{-\vec{r}}$ on the mass profile, signalling the vanishing of the mass term along some inversion symmetric curve. For concreteness, we consider the setup illustrated in Fig.~\ref{figure_for_sto} wherein the mass changes sign across the equator. It is known that such a domain wall hosts a chiral fermionic mode \citep{Jansen:1994ym}.
In order to gap out the chiral hinge mode, we induce topological orders denoted as $\mathcal A_{\mathrm{N}}$ and $\mathcal A_{\mathrm{S}}$ on the top/bottom halves of the surface. It is worth noting that the calculations below will be similar to the calculations in Ref.~\onlinecite{apoorv_2019}, albeit with an explicit treatment of crystalline symmetry.
3D inversion symmetry imposes  $\mathcal{A}_\mathrm{S}=\bar{\mathcal{A}}_\mathrm{N}$, where $\bar{\mathcal{A}}$ denotes the orientation-reversed version of $\mathcal{A}$. 
The topological order $\left[\text{Ising}\times \ms{U}(1)_{-8}\right]/\mathbb Z_{2}$ is a suitable choice for $\mathcal A_{\mathrm{N}}$.
This is the same topological order as the $\mathcal T$-Pfaffian, if we ignore TRS, therefore we refer to it as 2D $\mathcal T$-Pfaffian. 
The edge theory of $\mathcal A_{\mathrm{N}}$ contains a chiral Majorana mode and an anti-chiral compact boson mode.
These are the edge fields corresponding to the bulk $\text{Ising}$ and $\ms{U}(1)_{-8}$ topological orders respectively.
Concretely, the edge of $\mathcal{A}_\mathrm{N}$ and $\mathcal A_{\mathrm{S}}$ are described by the Lagrangians \cite{Bonderson_2013, Chen_2014}
\begin{align}
\mathcal{L}_{\partial \mathcal{A}_\mathrm{N}}=&\;-\frac{2}{4\pi}\partial_x \phi_\mathrm{N}(\partial_t-\partial_x)\phi_\mathrm{N}+i \gamma_\mathrm{N}(\partial_t+\partial_x)\gamma_\mathrm{N}, \nonumber \\
\mathcal{L}_{\partial \mathcal{A}_\mathrm{S}}=&\;-\frac{2}{4\pi}\partial_x \phi_\mathrm{S}(\partial_t-\partial_x)\phi_\mathrm{S}+i \gamma_\mathrm{S}(\partial_t+\partial_x)\gamma_\mathrm{S}.
\end{align}
where $\gamma_{\mathrm{N}}$ and $\gamma_{\mathrm{S}}$ are Majornana-Weyl modes while $\phi_{\mathrm{N}}$ and $\phi_{\mathrm{S}}$ are compact bosonic modes. The chiral hinge additionally contains a single Dirac mode contributed by the HOTI bulk  which can be described by the bosonized Lagrangian \cite{Wen_1990, Wen_1995, Senechal_2004}
\begin{equation}
\mathcal{L}_{0}=\frac{1}{4\pi}\partial_x \phi(\partial_t+\partial_x)\phi,
\label{eq:chiralmode_from_bulk}
\end{equation}
where $\phi$ is a compact boson. The combined Lagrangian describing the equatorial hinge is therefore given by  
\begin{align}
\mathcal{L}_{\text{Hinge}}=\mathcal{L}_{\partial \mathcal{A}_\mathrm{N}}+\mathcal{L}_{\partial \mathcal{A}_\mathrm{S}}+\mathcal{L}_{0}.   
\end{align}
The two Majorana modes can be combined into a Dirac mode which can be subsequently bosonized and written in terms of the compact boson $\phi_{\mathrm{M}}$ using 
\begin{equation}
\psi_{\mathrm{M}}\sim e^{i \phi_\mathrm{M}(x)} \sim e^{-i \frac{\pi}{4}}\gamma_\mathrm{N}+e^{i \frac{\pi}{4}}\gamma_\mathrm{S}, \label{bosonized:1}
\end{equation}
where we have suppressed the Klein factors for brevity \cite{von_Delft_1998}. The benefit of bosonizing the Majorana pair is that it allows for the description of the hinge in terms of the $K$-matrix Luttinger liquid formalism which is easier to work with. The hinge is then described by the Lagrangian
\begin{align}
\mathcal L_{\text{Hinge}}=\frac{1}{4\pi}\partial_{x}\Phi^{\ms{T}} K \partial_t \Phi -\frac{1}{4\pi}\partial_{x}\Phi^{\ms{T}}\partial_{x}\Phi,   
\label{eq:HOTI_hinge}
\end{align}
where $\Phi^{\ms{T}}=(\phi_{\mathrm{M}}, \phi_{\mathrm{N}},\phi_{\mathrm{S}},\phi)$, the $K=\text{diag}(1,-2,-2,1)$ and the charge vector $t^{\ms{T}}=(0,1,1,1)$. Our intention is to add to Eq.~\eqref{eq:HOTI_hinge} generic interactions represented by cosine terms that gap out all the hinge modes when driven to strong coupling  
\begin{align}
\delta \mathcal{L}&=\sum_{I=1,2}\delta\mathcal L_{I}=\sum_{I=1,2}\lambda_I(x)\cos[\Lambda_I^{\mathsf{T}}K\Phi-\alpha_{I}] 
\label{eq:gap_terms}
\end{align}
Apart from being simultaneously gappable (see App.~\ref{appendix:k-matrix} for details), the gapping vectors $\Lambda_{I}$ need to satisfy a number of symmetry criteria related to inversion, charge conservation and gauge symmetry derived from a $\mathbb Z_{2}$ redundancy in our description of the hinge. First, the inversion symmetry acts as $\mathcal I: \Phi(x)\mapsto \mathrm{I}\Phi(-x)$ where $\mathrm{I}=(-1)\oplus \sigma^{x}\oplus (+1)$. Second, in order to respect $\mathsf{U}(1)$ symmetry, we impose charge neutrality condition, namely we require that $\Lambda^{\mathsf{T}}t=0$. Finally, due to the fermionic nature of $\mathcal{A}_\mathrm{N}$ and $\mathcal{A}_\mathrm{S}$, an additional gauge symmetry $\mathbb{Z}^\mathrm{N}_2\times \mathbb{Z}^\mathrm{S}_2$ is imposed. The generators $\ms g_{\alpha}$ of $\mathbb Z_{2}^{\alpha}$ (where $\alpha=\mathrm{N},\mathrm{S}$) implement the transformation
\begin{align}
\ms g_{\mathrm{N},\mathrm{S}}&: \phi_{\mathrm{N},\mathrm{S}}\mapsto \phi_{\mathrm{N},\mathrm{S}}\pm\frac{\pi}{2},\quad  \gamma_{\mathrm{N},\mathrm{S}}\mapsto -\gamma_{\mathrm{N},\mathrm{S}}.
\end{align}
Since the fermionic operator $\Psi_{\alpha}\simeq\gamma_{\alpha} e^{2i\phi_{\alpha}}$ is invariant under $\mathbb Z_{2}^{\alpha}$, the gauge symmetry imposes that any admissible cosine term tunnels only local operators, that is, fermions or combinations thereof. Additionally, as a consequence of the $\mathbb Z_{2}^{\mathrm{N}}\times \mathbb Z_{2}^{\mathrm{S}}$ symmetry we need to fix the compactification $\phi_{\mathrm{N},\mathrm{S}}\sim \phi_{\mathrm{N},\mathrm{S}}+\pi$. 
Two cosine terms are required to open a gap for the combined hinge theory. The first gapping vector can be chosen to be  $\Lambda_1^{\mathsf{T}}=(0,-2,-2,4)$. Such a term is inversion-symmetric if $\lambda_1(x)=\lambda_1(-x)$ [see Eq.~\eqref{eq:gap_terms}] therefore $\lambda_{1}$ can be chosen to be constant and $\alpha=0$, i.e., 
\begin{align}
    \delta \mathcal L_{1}=\lambda_{1}\cos\left(4\phi_{\mathrm{N}}+4\phi_{\mathrm{S}}+4\phi\right).
    \label{eq:gap_term_1}
\end{align}
This gapping term also respects $\mathsf{U}(1)$ and $\mathbb{Z}^\mathrm{N}_2\times \mathbb{Z}^\mathrm{S}_2$ symmetry as can be checked explicitly. Upon adding Eq.~\eqref{eq:gap_term_1} to the original gapless hinge described by Eq.~\eqref{eq:HOTI_hinge}, the combination of fields $\langle \phi_{\mathrm{N}}+\phi_{\mathrm{S}}+\phi\rangle $ acquire a vacuum/groundstate expectation value, thereby breaking the $\mathbb{Z}^\mathrm{N}_2\times \mathbb{Z}^\mathrm{S}_2$ symmetry into a diagonal $\mathbb Z_{2}$ subgroup denoted as $\mathbb Z_{2}^{\text{diag}}$, generated by $\ms{g}_{\text{diag}}:=\ms g_{\mathrm{N}}\ms g_{\mathrm{S}}$ with the action
\begin{equation}
\ms{g}_{\text{diag}}: \phi_\mathrm{N}\mapsto \phi_\mathrm{N}+\frac{\pi}{2}, \phi_\mathrm{S}\mapsto \phi_\mathrm{S}-\frac{\pi}{2},\phi_{\mathrm{M}}\mapsto \phi_{\mathrm{M}}+\pi.
\end{equation}
The second gapping vector can be chosen as $\Lambda_2^{\mathsf{T}}=(2,1,-1,0)$. Since the two gapping vectors $\Lambda_{1,2}$ satisfy the Haldane criterion $\Lambda_i^{\mathsf{T}}K \Lambda_j=0$, the bosonic fields $\Lambda^{\ms{T}}K\Phi$ can simultaneously acquire a vacuum expectation value. The second gapping term $\delta \mathcal L_2$ is also charge neutral since $\Lambda_2^{\ms{T}}t=0$ as well as invariant under the residual $\mathbb Z_{2}^{\text{diag}}$ symmetry. Finally, the term is $\delta \mathcal L_{2}$ is inversion symmetric if $\alpha_{2}=n\pi$ and $\lambda_{2}(-x)=(-1)^{n}\lambda_{2}(x)$. By choosing $n\in \mathbb Z_{\text{even}}$, we can fix $\lambda_{2}$ to be constant everywhere. 

To summarize, we have shown that the two cosine terms corresponding to $\Lambda_{1,2}$ satisfy the symmetry requirements as well as the Haldane criteria. Therefore, they can be simultaneously driven to strong coupling thereby completely gapping out the hinge without breaking any symmetry. We note that an inversion symmetric configuration of $\mathcal A_{\mathrm{N}}$ and $\mathcal A_{\mathrm{S}}$ illustrated in Fig.~\ref{figure_for_sto} without the mode contributed from the bulk is clearly ingappable on the hinge as the modes on the hinge carry a non-vanishing chiral central charge. Therefore as a pattern of 2D inversion symmetric topological order, $\mathcal A_{\mathrm{N}}\oplus \mathcal A_{\mathrm{S}}$ is anomalous and cancels the higher-order anomaly coming from the bulk.  
\subsection{Class AII $+$ Inversion: HOTI with Helical Dirac Hinge Mode} \label{Sec:AII_HOTI}
\begin{figure}[tbh!]
\centering
\subfloat[Before pasting STO]{%
  \includegraphics[width=.475\linewidth]{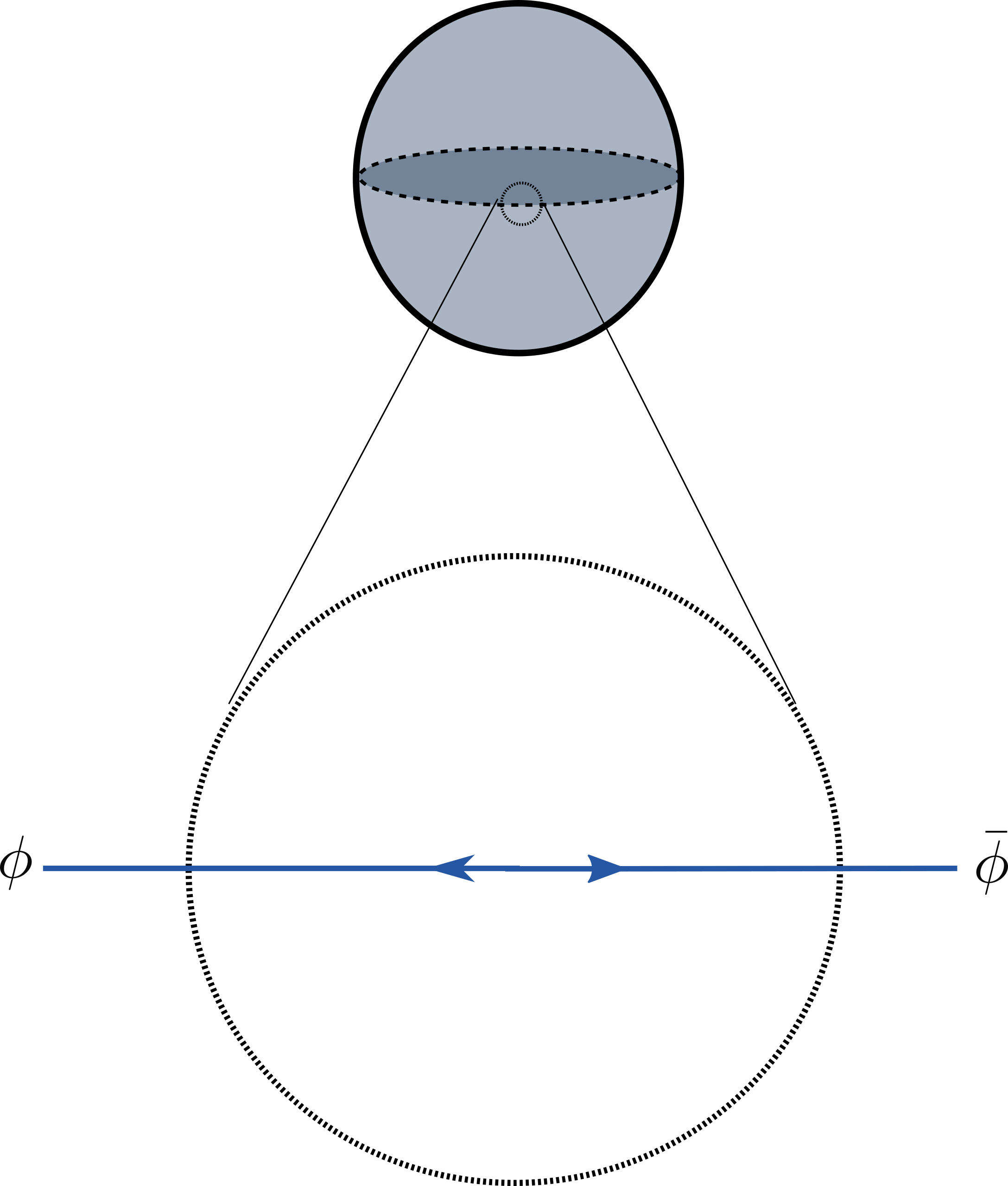}%
}\hfill
\subfloat[After pasting STO]{%
  \includegraphics[width=.49\linewidth]{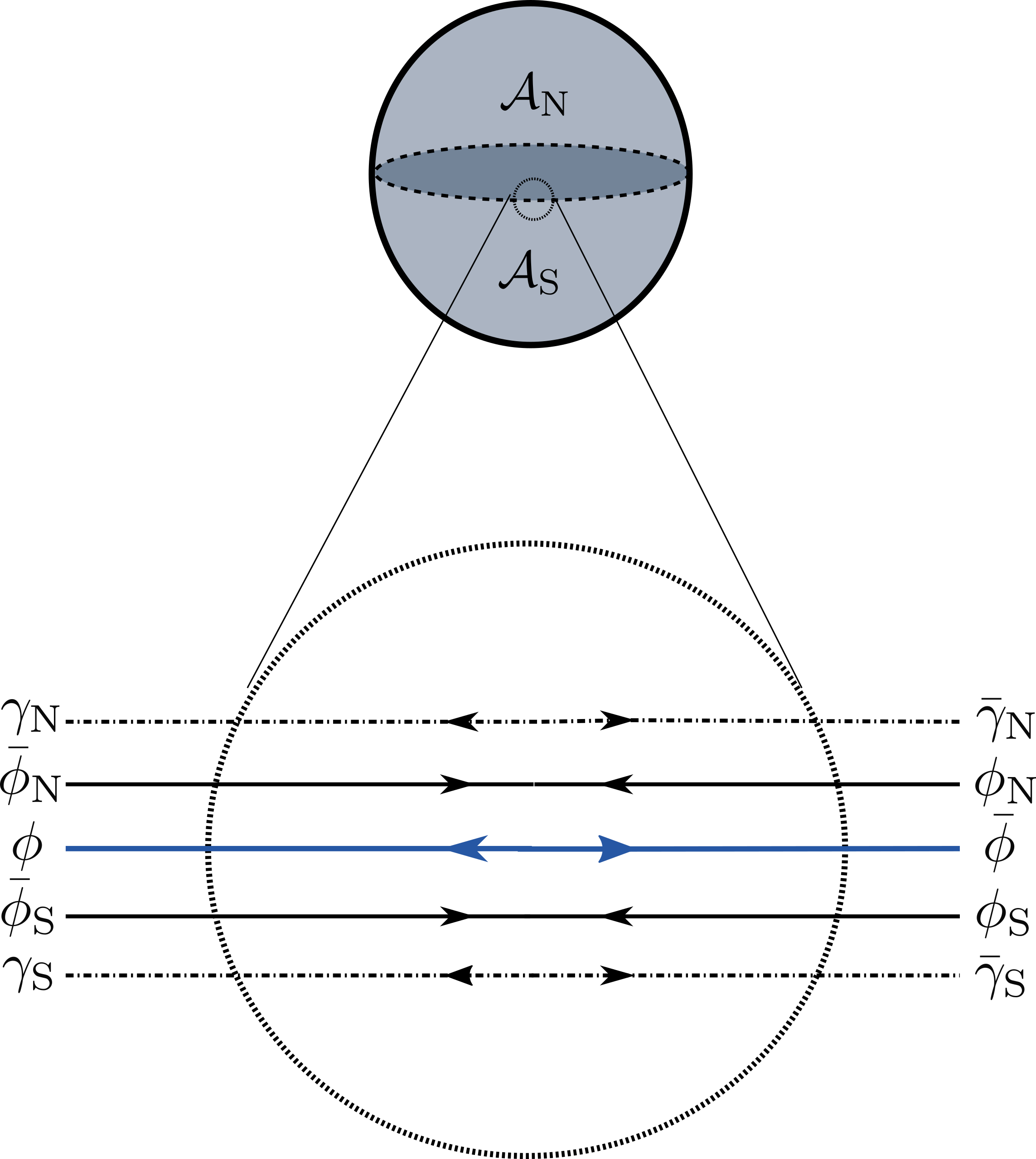}%
}
\caption{Surface topological order configuration for inversion and TRS protected higher-order topological insulator. }
\label{figure_for_sto:2}
\end{figure}
It was shown in Refs.~\onlinecite{khalaf_inversion, Khalaf_PRX} that TRS-invariant insulators (with $\mathcal T^{2}=-1$) enriched by additional inversion symmetry can support non-trivial second-order topology. On inversion symmetric open geometries, models within the non-trivial second-order phase host robust helical Dirac modes along an inversion symmetric hinge on the surface. The helical hinge modes are similar to those obtained on the edge of a quantum spin Hall insulator and form a Kramer's pair which is stable against interactions \cite{Kane-Mele-2005}. Here we show that these modes can be gapped out by inducing topological order on the surface. Before getting into the details of the surface topological order, we briefly review the free fermion model for the helical HOTI. The strategy is to start with a doubled model and subsequently add a perturbation that gaps out the surface leaving behind a robust hinge. In this case, the parent theory consists of two 3D topological insulators. We consider the same geometrical settings as in Fig.~\ref{figure_for_sto:2}. The surface Hamiltonian is given by
\begin{equation}
h_1(\vec{k},\vec{r})=\tau_0\otimes (\vec{k}\times \hat{n}_{\vec{r}})\cdot\vec{\sigma}.
\end{equation}
The inversion and TRS are represented by $\mathcal{I}=\tau_0\otimes (-\sigma_0)$ and $\mathcal{T}=\tau_0\otimes (i\sigma_y\mathcal{K})$. A surface mass term that respects TRS can be added to the Hamiltonian
\begin{equation}
    \delta h_1(\vec{r})=\tau_y\otimes m_{\vec{r}}(\hat{n}_{\vec{r}}\cdot \vec{\sigma}).
\end{equation}
Inversion symmetry demands that $\mathcal{I}\delta h_1(\vec{r})\mathcal{I}^{-1}=\delta h_1(-\vec{r})$ which further imposes the condition $m_{\vec{r}}=-m_{-\vec{r}}$,  signalling the vanishing of the mass term along some inversion symmetric curve. For reasons identical to the chiral HOTI case, this indicates the existence of gapless helical modes along the equator. TRS in the above construction acts within each flavor of the above model, i.e diagonally in the $\vec{\tau} $ space. We find it convenient to work with an equivalent description of the helical HOTI in which TRS acts by switching fermion flavors, such that the model can be thought of as a stacking of a chiral HOTI with its time-reversed copy. We consider the following Hamiltonian which is related to $h_1+\delta h_{1}$ by a unitary transformation
\begin{align}
h_2(\vec{k}, \vec{r})&=\tilde{\tau}_0 \otimes (\lambda_y\tilde{\sigma}_y+m_{\vec{r}}n_x\tilde{\sigma}_x+m_{\vec{r}}n_z\tilde{\sigma}_z)\nonumber \\
&+\tilde{\tau}_z \otimes (\lambda_x\tilde{\sigma}_x+\lambda_z\tilde{\sigma}_z+m_{\vec{r}}n_y\tilde{\sigma}_y), \label{hamiltonian:2}
\end{align}
where $\vec{\lambda}\equiv \vec{k}\times \hat{n}_{\vec{r}}$. TRS is represented by 
\begin{equation}
\mathcal{T}=i\tilde{\tau}_y\otimes \tilde{\sigma}_0 \mathcal{K}, \quad \ \mathcal T^{2}=-1. \label{trs:2}
\end{equation}
Upon performing an analysis similar to the one described above one is left with a surface which is gapped everywhere except an inversion symmetric hinge which hosts a pair of gapless helical modes $\left\{\psi, \bar{\psi}\right\}$ that form a Kramers doublet.

\noindent A natural candidate for an STO that can gap out the helical hinge mode is given by stacking the STO from the previous section and its time-reversed copy. It remains to be shown that this construction furnishes modes on the hinge that are robust by themselves, but when considered along with the helical modes contributed by the bulk lead to a completely gapped hinge. As illustrated in Fig.~\ref{figure_for_sto:2}, we set up the STO configuration by placing  $(\mathcal{A}_{\mathrm{N}},\mathcal{A}_{\mathrm{S}})$ on the northern/southern hemispheres of a spherical surface geometry. Here $\mathcal A_{\mathrm{N},\mathrm{S}}$ stand for the product topological orders consisting of 2D $\mathcal{T}$-Pfaffian topological orders and their time-reversed copies. The edge theory contains 8 modes. We divide these into bosonic modes $\Phi=(\phi_\mathrm{N}, \phi_\mathrm{S},\bar{\phi}_{\mathrm{N}},\bar{\phi}_{\mathrm{S}})^{\ms{T}}$ and fermionic modes $\Gamma=(\gamma_{\mathrm{N}},\gamma_{\mathrm{S}},\bar{\gamma}_\mathrm{N},\bar{\gamma}_\mathrm{S})^{\ms{T}}$. The action of TRS is encoded in the matrices $T_{\Phi}:=\sigma^{x}\otimes \mathbb 1_{2}$ and $T_{\Gamma}=i\sigma^{y}\otimes\mathbb 1_{2}$ such that under TRS 
\begin{align}
\mathcal T:&\; 
\begin{bmatrix}
\Phi \\
\Gamma
\end{bmatrix}
\longmapsto 
\begin{bmatrix}
T_{\Phi}\Phi \\
T_{\Gamma}\Gamma
\end{bmatrix}, \quad 
i \ \longmapsto -i.
\end{align}
As before we need to impose a gauge symmetry that ensures that the cosine terms only tunnel combinations of fields that are built from local fermionic operators. The full fermionic gauge symmetry group is $\mathbb Z_{2}^{4}=\prod_{\alpha}\mathbb{Z}^{\alpha}$ where $\alpha=\mathrm{N,S,\bar{N},\bar{S}}$. The generators of this group denoted as $\ms{g}_{\alpha}$ act as
\begin{align}
    \ms{g}_{\alpha}:&\; \begin{bmatrix}
    \gamma_{\alpha} \\
    \phi_{\alpha}
    \end{bmatrix}
    \longmapsto
    \begin{bmatrix}
    -\gamma_{\alpha} \\
    \phi_{\alpha}
    \end{bmatrix}
    +
   s_{\alpha} \begin{bmatrix}
    0 \\
    \frac{\pi}{2}
    \end{bmatrix},
\label{eq:fourgauge}
\end{align}
where $s_{\alpha}=-1$ for $\alpha=\mathrm{S},\mathrm{S}_{T}$ and $+1$ otherwise. Inversion squares to +1 and simply maps the fields on the northern hemisphere to their counterparts on the southern hemisphere and vice versa. We now proceed to gap out the edge modes. Firstly we combine the Majorana fermions into Dirac fermions
\begin{align}
    \psi_\mathrm{M}&\sim e^{i\phi_\mathrm{M}}\sim e^{-i\frac{\pi}{4}}\gamma_\mathrm{N}+e^{i\frac{\pi}{4}}\gamma_\mathrm{S}, \nonumber \\ \bar{\psi}_\mathrm{M}&\sim e^{-i\bar{\phi}_\mathrm{M}}\sim e^{i\frac{\pi}{4}}\bar{\gamma}_\mathrm{N}+e^{-i\frac{\pi}{4}}\bar{\gamma}_\mathrm{S}.
\end{align}
Since the Majorana fermions are themselves Kramers pairs, the action of TRS can be deduced as
\begin{align}
\mathcal T:&\; 
\begin{bmatrix}
\phi_\mathrm{M} \\
\bar{\phi}_\mathrm{M}
\end{bmatrix}
\longmapsto 
\begin{bmatrix}
\bar{\phi}_\mathrm{M} \\
\phi_\mathrm{M}+\pi
\end{bmatrix}.
\end{align}
The edge is now effectively described by the following $K$-matrix and charge vector $t$
\begin{align}
    K&=\mathrm{diag}(-1,2,2,-1,1,-2,-2,1),\nonumber \\
    t&=(0,1,1,1,0,1,1,1)^{\mathsf{T}},
\end{align}
in the basis $(\phi_\mathrm{M},\phi_\mathrm{N}, \phi_\mathrm{S},\phi,\bar{\phi}_\mathrm{M},\bar{\phi}_{\mathrm{N}},\bar{\phi}_{\mathrm{S}},\bar{\phi})^{\ms{T}}$. Consider the gapping terms  
\begin{align}
    \hspace{-7pt}\delta \mathcal{L}&=\cos[4\phi_\mathrm{N}+4\phi_\mathrm{S}+4\phi]+\cos[2\phi_\mathrm{N}-2\phi_\mathrm{S}-2\phi_\mathrm{M}]\nonumber \\
    &+\cos[4\bar{\phi}_\mathrm{N}+4\bar{\phi}_\mathrm{S}+4\bar{\phi}]+\cos[2\bar{\phi}_\mathrm{N}-2\bar{\phi}_\mathrm{S}-2\bar{\phi}_\mathrm{M}].
\end{align}
Note that this expression is basically the gapping term for an inversion symmetric HOTI plus its time-reversed copy. Thus, we only need to check whether the above expression breaks TRS.  Clearly, it does not break TRS explicitly; however, since both $\langle \bar{\phi}_\mathrm{N}+\bar{\phi}_\mathrm{S}+\bar{\phi}\rangle $ and $\langle\bar{\phi}_\mathrm{N}-\bar{\phi}_\mathrm{S}-\bar{\phi}_\mathrm{M} \rangle$ transform to their TRS copies with extra $\pi$ phase, naively it seems like TRS is broken spontaneously. We note that the gauge group is now broken to $\mathbb{Z}^{\mathrm{diag}}_2\times \bar{\mathbb{Z}}^{\mathrm{diag}}_2$, where
\begin{align}
\ms{g}_{\text{diag}}&: \phi_\mathrm{N}\mapsto \phi_\mathrm{N}+\frac{\pi}{2}, \phi_\mathrm{S}\mapsto \phi_\mathrm{S}+\frac{\pi}{2},\phi_{\mathrm{M}}\mapsto \phi_{\mathrm{M}}+\pi,\nonumber \\
\bar{\ms{g}}_{\text{diag}}&: \bar{\phi}_\mathrm{N}\mapsto \bar{\phi}_\mathrm{N}+\frac{\pi}{2}, \bar{\phi}_\mathrm{S}\mapsto \bar{\phi}_\mathrm{S}+\frac{\pi}{2},\bar{\phi}_{\mathrm{M}}\mapsto \bar{\phi}_{\mathrm{M}}+\pi.
\end{align}
We can see that $\langle \phi_\mathrm{N}+\phi_\mathrm{S}+\phi \rangle\sim \langle \phi_\mathrm{N}+\phi_\mathrm{S}+\phi \rangle+\pi$, $\langle \phi_\mathrm{N}-\phi_\mathrm{S}-\phi_\mathrm{M} \rangle\sim \langle \phi_\mathrm{N}-\phi_\mathrm{S}-\phi_\mathrm{M} \rangle+\pi$ as they are related to each other by a gauge transformation. Therefore TRS is not broken spontaneously either. Finally, we emphasize that without  inversion symmetry, the above STO is non-anomalous as we can paste a copy of a quantum spin Hall liquid on, e.g., the southern hemisphere, with its edge modes residing on the equator and gap out the helical modes contributed by the STO without invoking the bulk.

\subsection{Class D $+$ Inversion: HOTSC with Chiral Majorana Hinge Mode} \label{Sec:ClassD_HOTSC}
We briefly review the free fermion model for the chiral HOTSC in class D. The strategy to construct a second-order phase is to start with a class DIII topological superconductor and add a TRS-breaking perturbation that gaps out the surface leaving behind a robust hinge protected by inversion symmetry. The surface Hamiltonian is given by \cite{khalaf_inversion}
\begin{equation}
h(\vec{k},\vec{r})=-(\vec{k}\times \hat{n}_{\vec{r}})\cdot\vec{\sigma},
\end{equation}
The inversion, time reversal and particle-hole symmetries are generated by $\mathcal{I}=-\sigma_0$, $\mathcal{T}=i\sigma_y \mathcal K$ and $\mathcal P= -(\hat{n}_{\vec{r}}\cdot\vec{\sigma})\sigma_{y}\mathcal{K}$. The surface can be deformed by the mass term
\begin{equation}
\delta h(\vec{r})=m_{\vec{r}}(\hat{n}_{\vec{r}}\cdot \vec{\sigma}),
\end{equation}
that breaks TRS. Inversion symmetry demands that $\mathcal{I}\delta h(\vec{r})\mathcal{I}^{-1}=\delta h(-\vec{r})$ which further imposes the condition $m_{\vec{r}}=-m_{-\vec{r}}$, signalling the vanishing of the mass term along some inversion symmetric curve which hosts gapless chiral modes. Due to the additional particle-hole symmetry as compared with the class A chiral HOTI, these are Majorana as opposed to Dirac modes. 
\begin{figure}[t]
\includegraphics[width=0.45\textwidth]{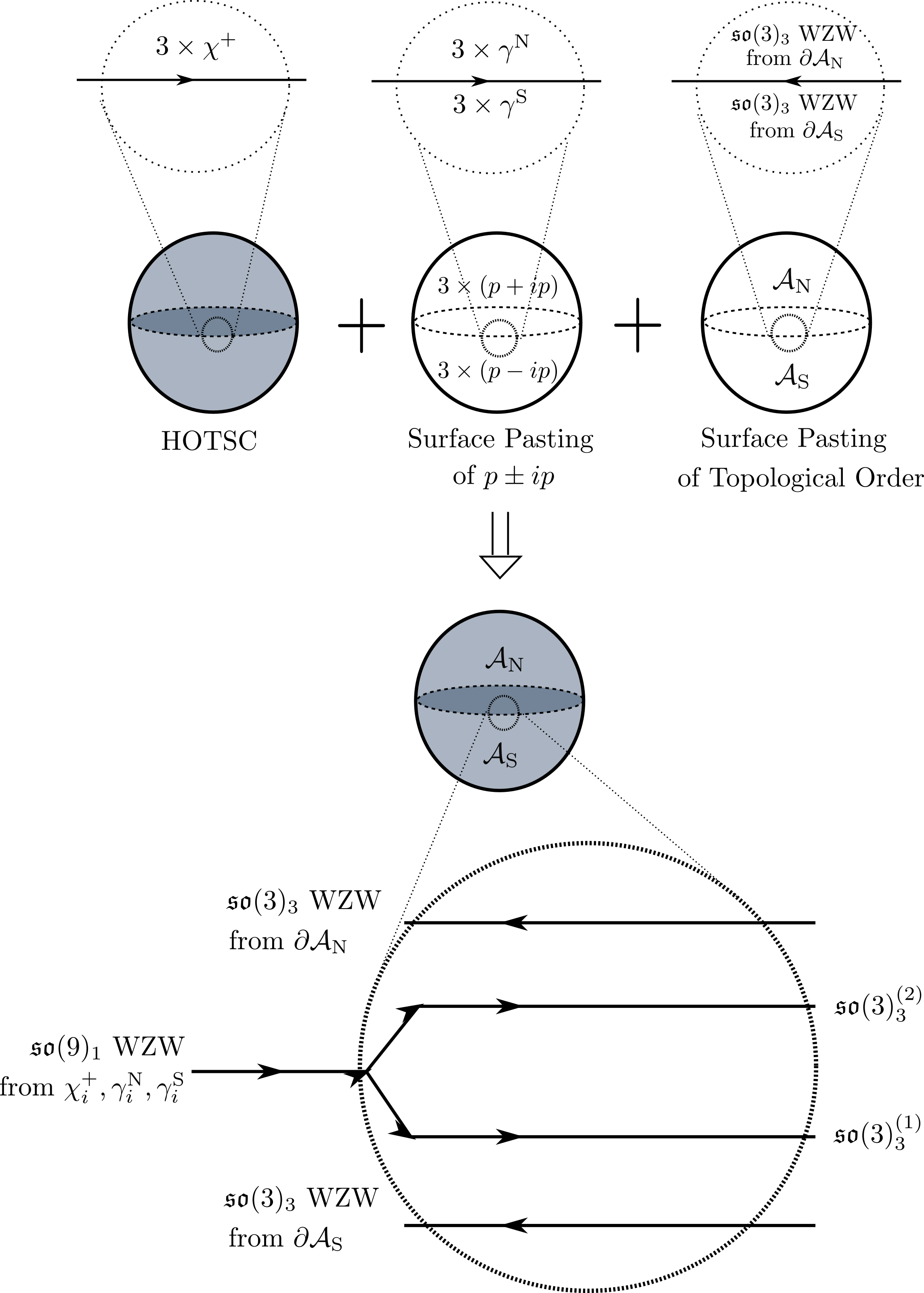}
\caption{An illustration of the procedure used to gap out the $(3,0)$ configuration of chiral Majorana modes $\chi_{i}$ with $i=1,2,3$. We introduce $\gamma^{N/S}_i$ Majoranas by pasting $p\pm ip$ superconductors on the surface and end up with a total of nine Majorana modes on the hinge. These are subsequently gapped out by introducing the STO $\mathcal A_{N,S}$ on the northern and southern hemispheres respectively. The nine Majorana modes are described by the $\ms{SO}(9)_1$ WZW which splits into two copies of $\ms{SO}(3)_3$ theories that gap out upon coupling to the edge modes provided by the STOs.}
\label{STO_Chiral_HOTSC}
\end{figure}
 Before turning to the surface topological order, we first inspect the stability of the chiral hinge modes to inversion symmetric surface pasting of $p\pm ip$ superconductors.
  Let us consider the situation where there are $N_{\pm}$ chiral co-propagating Majorana hinge modes denoted as $\chi^{\pm}_{i}$ with $i=1,\dots,N_{\pm}$ that transform under inversion as
\begin{align}
    \mathcal I:
    \begin{bmatrix}
    \chi_{i}^{\pm}(\theta)\\
    \chi_{i}^{\pm}(\theta+\pi)
    \end{bmatrix}\mapsto 
    \begin{bmatrix}
    \pm i\chi_{i}^{\pm}(\theta+\pi)\\
    \mp i\chi_{i}^{\pm}(\theta)
    \end{bmatrix},
\end{align}
where $\theta$ is introduced to parameterize the equator on which the Majoranas are propagating. The above symmetry action can be derived from the bulk symmetry using a recursive Jackiw-Rebbi procedure (see App.~\ref{App:Jackiw_Rebbi}) \citep{jackiw_rebbi, minghao_notes} and satisfies the basis-invariant relations $\{\mathcal{I}, \mathcal{P}\}=0$, $\left[\mathcal I,\mathcal T\right]=0$ and $\mathcal I^{2}=1$. we can always paste a $p+ ip$ and $p- ip$ topological superconductor on the northern and southern hemispheres respectively, which contribute a pair of chiral hinge modes denoted as $\gamma^{\mathrm{N},\mathrm{S}}$. The inversion action on these modes is represented as
\begin{align}
    \mathcal I:
    \begin{bmatrix}
    \gamma^{\mathrm{N}}(\theta)\\
    \gamma^{\mathrm{N}}(\theta+\pi)\\
    \gamma^{\mathrm{S}}(\theta)\\
    \gamma^{\mathrm{S}}(\theta+\pi)
    \end{bmatrix}\mapsto 
    \begin{bmatrix}
    i\gamma^{\mathrm{S}}(\theta+\pi)\\
    -i\gamma^{\mathrm{S}}(\theta)\\
    i\gamma^{\mathrm{N}}(\theta+\pi)\\
    -i\gamma^{\mathrm{N}}(\theta)
    \end{bmatrix}.
\end{align}
Consider the linear combinations $\gamma^{\pm}=(\gamma^{\mathrm{N}}\pm \gamma^{\mathrm{S}})/\sqrt{2}$ that transform under inversion as
\begin{align}
    \mathcal I:
    \begin{bmatrix}
    \gamma^{\pm}(\theta)\\
    \gamma^{\pm}(\theta+\pi)
    \end{bmatrix}\mapsto 
    \begin{bmatrix}
    \pm i\gamma^{\pm}(\theta+\pi)\\
    \mp i\gamma^{\pm}(\theta)
    \end{bmatrix}.
\end{align}
Henceforth we denote left/right-moving modes with/without an overbar. The configuration $(N_{+},N_{-})$ with net $N_{+}+N_{-}$ right movers can always be transformed to $(N_{+}-1,N_{-}-1)$ by surface pasting. Therefore we have the equivalence relation 
\begin{align}
    (N_{+},N_{-})\sim (N_{+}-n,N_{-}-n),
    \label{eq:hinge_config}
\end{align}
where $n\in \mathbb Z$. Consequently, we can always transform a configuration $(N_{+},N_{-})$ into a configuration with all positive parity modes $(N_{+}-N
_{-},0)$.
For this reason we will only need to consider the stability of such modes under surface pasting of topologically ordered phases. 
The classification of inversion symmetry-protected higher-order phases in class D is given by the group $\mathbb Z_{4}$ which can be indexed by $(N_+,0)$ (See Sec.~\ref{Sec:3rdOrder} for details).
For the present discussion it will suffice to construct the STO for the generator of $\mathbb Z_{4}$ which may be treated as $(3,0)$.  
Since an odd number of Majorana fermions cannot be tamed by Abelian bosonization, the $K$-matrix approach we previously employed must be abandoned.
Instead, we use non-Abelian bosonization to approach the problem.
We remark here that the method in this section is similar in spirit to Ref.~\onlinecite{Meng_2018,teo_majorana}, however, adapted to inversion symmetry. 

We consider the HOTSC to have a spherical geometry as illustrated in Fig.~\ref{STO_Chiral_HOTSC} with three chiral majorana modes $\chi_{i}^{+}$ on the inversion-symmetric equator. In order to show that the chiral hinge can be gapped, we find it convenient to proceed in two steps. First we add additional degrees of freedom on the hinge by a purely surface pasting of $p\pm ip$ superconductors that preserves the inversion symmetry. 
In the second step, we induce inversion-symmetric topological order on the surface to gap out the combined hinge modes contributed by the $p\pm ip$ superconductors and the bulk higher-order superconductor.
We begin by adding three copies of a $p+ip$ superconductor on the northern hemisphere and three copies of a $p-ip$ superconductor on the southern hemisphere.  
As a result, we end up with $6$ additional Majorana modes on the equator which we label as $\{\gamma^\mathrm{N}_i,\gamma^\mathrm{S}_i \}$, with $i=1,\dots, 3$.
The hinge is  described by the $\mathfrak{so}(9)_1$ WZW theory \citep{cft_fran} 
\begin{align}
    S=\int\mathrm{d}t\mathrm{d}\theta i{\Psi}^{\ms{T}}(
    \partial_{t}-\partial_{\theta}){\Psi},
    \label{eq:so(9)_1_wzw}
\end{align}
where we have introduced a 9 component Majorana spinor field $\Psi$. The operator product expansion (OPE) of the Majorana operators satisfies the standard relations
\begin{align}
    \chi_{i}^{+}(z)\chi_{j}^{\mathrm{N}}(w)\sim&\; \frac{\delta_{ij}}{z-w}+\dots, \nonumber \\
    \gamma_{i}^{\mathrm{N}}(z)\gamma_{j}^{\mathrm{N}}(w)\sim&\; \frac{\delta_{ij}}{z-w}+\dots, \nonumber \\
    \gamma_{i}^{\mathrm{S}}(z)\gamma_{j}^{\mathrm{S}}(w)\sim&\; \frac{\delta_{ij}}{z-w}+\dots.
\end{align}
We introduce $\mathfrak{so}(9)_{1}$ currents which can be expressed as fermion biliears, 
\begin{align}
    \mathcal J^{\ms{A}}(z)=&\; \frac{i}{2}{\Psi^{\dagger}}(z)\Sigma^{\ms{A}}{\Psi}(z),
    \label{eq:current_defn}
\end{align}
where $\ms{A}$ is a Lie-algebra index, $\Sigma^{\ms{A}}$ are the generators of the $\mathfrak{so}(9)$ lie algebra and $z$ are holomorphic coordinates defined as $z=\theta+i t$ on the hinge. The currents $\left\{\mathcal J^{\ms{A}} \right\}_{\ms{A}=1,\dots \text{dim}(\mathfrak{so}(9))}$ satisfy the OPE
\begin{align}
    \mathcal{J}^{\ms{A}}(z)    \mathcal{J}^{\ms{B}}(w)\sim \frac{\delta^{\ms{AB}}}{(z-w)^{2}}+ \frac{i f^{AB}_{C}\mathcal J^{C}(w)}{z-w} + \dots,
\end{align}
where $f^{\ms{AB}}_{\ms{C}}$ are the structure constants for $\mathfrak{so}(9)$. The action of inversion on the different Majorana operators is as follows
\begin{align}
    \mathcal I:&\; \chi^{+}_i(\theta) \longrightarrow i\chi^{+}_{i}(\theta+\pi), \nonumber \\
    \mathcal I :&\; \gamma_{i}^{\mathrm{N}}(\theta)\longrightarrow i\gamma_{i}^{\mathrm{S}}(\theta+\pi), \nonumber \\
     \mathcal I:&\; \gamma_{i}^{\mathrm{S}}(\theta)\longrightarrow i\gamma_{i}^{\mathrm{N}}(\theta+\pi).
\end{align}
In order to construct the surface topological order that can absorb the Majorana hinge modes, it is useful to work with an embedding of $ \mathfrak{so}(3)^{(1)}_{3}\times \mathfrak{so}(3)^{(2)}_3\subset \mathfrak{so}(9)_{1} $. Since inversion symmetry is an essential part of our setup, we need to be careful about its action on the various embedded components. We work with a choice of embedding such that the two copies of $\mathfrak{so}(3)_3$ are swapped under the action of inversion. Let us index the components of the spinor $\Psi$ by a tuple $(i,j)$ where $i,j=1,2,3$. We define the different components such that they have the following simple transformation rule under inversion
\begin{align}
    \mathcal I:&\Psi_{(i,j)}(\theta) \longmapsto i\Psi_{(j,i)}(\theta+\pi)
\end{align}
with
\begin{align}
\Psi_{(i,i)}=&\; \chi^{+}_{i}, \nonumber \\
\Psi_{(2,3)}=&\; \gamma_{1}^{\mathrm{N}}, \quad  \Psi_{(3,2)}= \gamma_{1}^{\mathrm{S}}, \nonumber \\
\Psi_{(3,1)}=&\; \gamma_{2}^{\mathrm{N}}, \quad  \Psi_{(1,3)}= \gamma_{2}^{\mathrm{S}}, \nonumber \\
\Psi_{(1,2)}=&\; \gamma_{3}^{\mathrm{N}}, \quad  \Psi_{(2,1)}= \gamma_{3}^{\mathrm{S}}.  
\label{tupleassignchiral}    
\end{align}  
In order to construct the $\mathfrak{so}(3)_3$ current operators, consider the matrices defined as  $\sigma^{\ms{a},1}:=\mathrm{L}^{\ms{a}}\otimes \mathrm{Id}_{3}$ and $\sigma^{\ms{a},2}:=\mathrm{Id}_{3}\otimes \mathrm{L}^{\ms{a}}$, $\ms{a}=1,2,3$, where $\mathrm{L}^{\ms{a}}$ are the generators of $\mathfrak{so}(3)$ in the fundamental representation. These matrices generate two decoupled $\mathfrak{so}(3)$ algebras
\begin{align}
    \left[\sigma^{\ms{a},\kappa},\sigma^{\ms{b},\kappa'}\right]=\delta^{\kappa\kappa'}f^{\ms{ab}}_{\ms{c}}\sigma^{\ms{c},\kappa}.
    \label{Eq:embedded_matrices}
\end{align}
Using this decomposition we define the following $\mathfrak{so}(3)_{3}\times \mathfrak{so}(3)_{3}$ currents
\begin{equation}
\mathcal J^{\ms{a},\kappa}=\frac{i}{2}\Psi^{\dag}\sigma^{\ms{a},\kappa}\Psi,
\end{equation}
which explicitly take the form
\begin{align}
    \mathcal J^{1,1}=&\; \frac{i}{2}\left[\left(\gamma^{\mathrm{S}}_{3}\right)^{(\dag)}\gamma^{\mathrm{N}}_{2}+\left(\chi_2^{+} \right)^{(\dag)}\gamma_{1}^{\mathrm{S}}+\left(\gamma_{1}^{\mathrm{N}}\right)^{(\dag)}\chi_{3}^{+}\right]+\text{h.c.}, \nonumber \\
    \mathcal J^{1,2}=&\; \frac{i}{2}\left[\left(\gamma^{\mathrm{N}}_{3}\right)^{(\dag)}\gamma^{\mathrm{S}}_{2}+\left(\chi_{2}^{+}\right)^{(\dag)}\gamma_{1}^{\mathrm{N}}+\left(\gamma_{1}^{\mathrm{S}}\right)^{(\dag)}\chi_{3}^{+}\right]+\text{h.c.}, \nonumber \\
    \mathcal J^{2,1}=&\; \frac{i}{2}\left[\left(\chi_{1}^{+}\right)^{(\dag)}\gamma^{\mathrm{N}}_{2}+\left(\gamma_{3}^{\mathrm{N}}\right)^{(\dag)}\gamma_{1}^{\mathrm{S}}+\left(\gamma_{2}^{\mathrm{S}}\right)^{(\dag)}\gamma_{3}^{+}\right]+\text{h.c.}, \nonumber \\
            \mathcal J^{2,2}=&\; \frac{i}{2}\left[\left(\chi_{1}^{+}\right)^{(\dag)}\gamma^{\mathrm{S}}_{2}+\left(\gamma_{3}^{\mathrm{S}}\right)^{(\dag)}\gamma_{1}^{\mathrm{N}}+\left(\gamma_{2}^{\mathrm{N}}\right)^{(\dag)}\chi_{3}^{+}\right]+\text{h.c.}, \nonumber \\
    \mathcal J^{3,1}=&\; \frac{i}{2}\left[\left(\chi_{1}^{+}\right)^{(\dag)}\gamma^{\mathrm{S}}_{3}+\left(\gamma_{3}^{\mathrm{N}}\right)^{(\dag)}\chi_{2}^{+}+\left(\gamma_{2}^{\mathrm{S}}\right)^{(\dag)}\gamma_{1}^{\mathrm{N}}\right]+\text{h.c.}, \nonumber \\
    \mathcal J^{3,2}=&\; \frac{i}{2}\left[\left(\chi_{1}^{+}\right)^{(\dag)}\gamma^{\mathrm{N}}_{3}+\left(\gamma_{3}^{\mathrm{S}}\right)^{(\dag)}\chi_{2}^{+}+\left(\gamma_{2}^{\mathrm{N}}\right)^{(\dag)}\gamma_{1}^{\mathrm{S}}\right]+\text{h.c}.. \label{currentschiralhinge}
\end{align}
The reason why we write $(\dag)$ on the Majorana operator is to remind ourselves of the subtlety related to the imaginary action of inversion, e.g., if $\mathcal I: \chi^{+}_i(\theta) \mapsto i\chi^{+}_{i}(\theta+\pi)$, then $\mathcal I: (\chi^{+}_i(\theta))^{(\dag)} \mapsto (\chi^{+}_{i}(\theta+\pi))^{(\dag)}(-i)$. We can verify that inversion acts as $\mathcal{I}: \mathcal{J}^{\ms{a},1}\leftrightarrow\mathcal{J}^{\ms{a},2}$ on the $\mathfrak{so}(3)$ currents. From the standard OPE for Majorana operators, we can extract the OPE for the $\mathfrak{so}(3)_{3}$ currents, and verify that the level is indeed 3 (see App.~\ref{appendix:wzw}). The stress tensor decomposes as 
\begin{equation}
T_{\mathfrak{so}(9)_1}=T_{\mathfrak{so}(3)^{(1)}_3}+T_{\mathfrak{so}(3)^{(2)}_3},
\end{equation}
which means that the chiral central charges of the embedded sectors add up to give the chiral central charge of the $\mathfrak{so}(9)_1$ WZW theory. Having formulated the hinge modes as two copies of $\mathfrak{so}(3)_3$, it is a straightforward task to gap them out by adding surface topological order. We introduce $\mathcal A_{\mathrm{N}}=\overline{\mathcal A_{\mathrm{S}}}=\overline{\ms{SO}(3)}_{3}$ whose edge conformal field theories and corresponding current operators we denote as $\overline{\mathfrak{so}(3)}_{3,\mathrm{N/S}}$ and $\bar{\mathcal J}^{\ms{a}}_{\mathrm{N/S}}$, respectively \cite{Meng_2018}. Under inversion the currents transform as
\begin{equation}
\mathcal{I}: \bar{\mathcal J}^{\ms{a}}_{\mathrm{N}}(\theta)\longmapsto \bar{\mathcal J}^{\ms{a}}_{\mathrm{S}}(\theta+\pi).
\end{equation}
The hinge modes $\mathcal{J}^{\ms{a},\kappa}$ and the edge modes of the surface topological order $\bar{\mathcal{J}}^{\ms{a}}_{\mathrm{N/S}}$ can together be gapped out upon adding the gapping term
\begin{equation}
\delta \mathcal{L}=\lambda(\theta)\sum_{\ms{a}=1}^{3}\Big[\bar{\mathcal J}^{\ms{a}}_\mathrm{N}(\theta)\mathcal J^{\ms{a},1}(\theta)+\bar{\mathcal J}^{\ms{a}}_\mathrm{S}(\theta)\mathcal J^{\ms{a},2}(\theta)\Big],
\end{equation}
which is inversion symmetric if $\lambda(\theta)=\lambda(\theta+\pi)$. Therefore, we can choose $\lambda$ to be constant. To summarize, we have shown that the hinge modes $(3,0)$ can first be mapped to $(3,6)$ by purely surface pasting of $p\pm ip$ superconductors. Thereafter, two copies of $\mathfrak{so}(3)_3$ can be embedded in the $(3,6)$ configuration which can be gapped out by a surface pasting of $\mathsf{SO}(3)_3$ topological order.

\subsection{Class DIII $+$ Inversion: HOTSC with Helical Majorana Hinge Modes}
\label{sec:STO_DIII}
\noindent Class DIII superconductors enriched by inversion symmetry support non-trivial second and third-order topological phases which host robust helical modes and Majorana Kramers zero modes on inversion symmetric loci on the surface \cite{khalaf_inversion}. The helical hinge modes are similar to those obtained on the edge of a 2D TRS invariant topological superconductor, i.e they contain a Majorana Kramers' pair of counter-propagating modes. Here we show that these modes can be gapped out if we allow for the possibility of surface topological order. 

First we briefly review the free-fermion model for the helical HOTSC. We can start with two class DIII topological superconductors with opposite topological index and add symmetry-respecting perturbations that gap out the surface, leaving behind a robust hinge. The surface Hamiltonian prior to adding such a mass term is given by \begin{equation}
h(\vec{k},\vec{r})=-\rho_z\otimes (\vec{k}\times \hat{n}_{\vec{r}})\cdot\vec{\sigma}, \label{eq:2ndorderdiii}
\end{equation}
where $\rho_{\mu}$ and $\sigma_{\mu}$ are the Pauli matrices in the orbital and spin space respectively. The inversion, time reversal and particle-hole symmetries are generated by $\mathcal{I}=-\rho_z\otimes \sigma_0$,  $\mathcal{T}=\rho_0\otimes i\sigma_y \mathcal {K}$ and $\mathcal P= -\rho_{z}\otimes (\hat{n}_{\vec{r}}\cdot \vec{\sigma})\sigma_{y}\mathcal{K}$. To perturb the Hamiltonian, we add the following mass term that respects the symmetries
\begin{equation}
\delta h(\vec{r})=m_{\vec{r}}\rho_x\otimes \sigma_0.\label{eq:mass2ndorderdiii}
\end{equation}
Inversion symmetry demands that $\mathcal{I}\delta h(\vec{r})\mathcal{I}^{-1}=\delta h(-\vec{r})$ which further imposes the condition $m_{\vec{r}}=-m_{-\vec{r}}$, signalling the vanishing of the mass term along some inversion symmetric curve which we choose to be at the equator. After Jackiw-Rebbi projection, we may verify that a pair of helical Majorana modes resides at the hinge. \par
The classification of inversion symmetry enriched higher-order phases in class DIII is given by the group $\mathbb Z_{4}$ which can be indexed by the number of Majorana helical hinge modes modulo 4 (See Sec.~\ref{Sec:3rdOrder} for details).
For the present discussion it will suffice to construct the STO for the generator of $\mathbb Z_{4}$ which may be treated as $(3,0)$. 
We first show that an odd number of helical modes is stable to weak interaction, and they are only unstable to adding topological order on the surface. 

\subsubsection{Stability of odd number of helical hinge modes}
For concreteness, let us begin with the setup with three pairs of helical Majorana modes on the hinge. The Lagrangian density for the Majorana modes can be written as
\begin{align}
    \mathcal L=\sum_{j=1}^{2n+1}\Big[i\chi_j(
    \partial_{t}-\partial_{x})\chi_{j}+i\bar{\chi}_j(
    \partial_{t}+\partial_{x})\bar{\chi}_{j}\Big],
    \label{eq:action_so(2n+1)WZW}
\end{align}
with $n=1$ which is the non-chiral $\mathfrak{so}(3)_{1}$ WZW theory. For simplicity, we drop the explicit hermitian conjugation from the equations in this subsection as they do not play a role unless we are dealing with a imaginary symmetry representation such as inversion in Eq.~\eqref{eq:inv_repn_maj}. Here we are only interested in the stability of pairs of helical-modes under TRS which has a real representation. The holomorphic and anti-holomorphic currents that generate the $\mathfrak{so}(3)$ current algebra are
\begin{align}
\mathcal J^{\ms{a}}=&\; \frac{i}{2}\chi_j \mathrm{L}^{\ms{a}}_{jk} \chi_k=\frac{i}{2}\epsilon^{\ms{a}jk}\chi_{j}\chi_{k} \nonumber \\
\bar{\mathcal J}^{\ms{a}}=&\; \frac{i}{2}\bar{\chi}_j \mathrm{L}^{\ms{a}}_{jk} \bar{\chi}_k=\frac{i}{2}\epsilon^{\ms{a}jk}\bar{\chi}_{j}\bar{\chi}_{k} ,
\end{align}
where $\ms{a}=1,\dots,\text{dim}(\mathfrak{so}(3))$, and as before $\mathrm{L}^{\ms{a}}$ are the generators for the $\mathfrak{so}(3)$ Lie algebra. The model is TRS invariant with the TRS action given by Eq.~\eqref{eq:TRS_Majorana} for each pair $(\chi_{j},\bar{\chi}_{j})$. We are interested in the stability of this model to TRS invariant perturbations. More precisely, whether the theory can be completely gapped out without breaking TRS. At the quadratic level we can add the following terms to the Hamiltonian
\begin{align}
\delta H =&\; \sum_{j,k,l}im_{j}\epsilon^{jkl}(\chi_k\bar{\chi}_l+\bar{\chi}_k\chi_l)+\sum_{j}i\widetilde{m}_j\chi_j\bar{\chi}_j \nonumber \\
=&\; \sum_{j}\left[ m_{j}\mathcal O^{j} + i\widetilde{m}_{j}\chi_{j}\bar{\chi}_{j}\right],
\end{align}
where, in the second second line we have defined the fermion bilinear $\mathcal O^{j}=i\epsilon^{jkl}(\chi_{k}\bar{\chi}_{l}
+\bar{\chi}_{k }\chi_{l})$. TRS imposes that $\widetilde{m}_{j}=-\widetilde{m}_{j}=0$, while there are no such constraints on $m_{j}$. The operators $\mathcal O^{j}$ satisfy the algebra
\begin{align}
    \left[\mathcal O^{j},\mathcal O^{k}\right]=4(\chi_{j}{\chi}_{k}+\bar{\chi}_{j}\bar{\chi}_{k}),
    \label{eq:non_abelian_haldane}
\end{align}
which suggests that these operators cannot condense/acquire a ground state expectation value simultaneously. This may pose an obstruction to symmetrically gapping out the theory. Since the model is quadratic, we simply diagonalize the Hamiltonian and check whether this is the case. The full Hamiltonian reads
\begin{align}
H=&\;\sum_j\int\mathrm{d}x\Big\{ iv \left(\chi_j\partial_x \chi_j-\bar{\chi}_j\partial_x \bar{\chi}_j\right)+\sum_{j}im_{j}\mathcal O^{j}\Big\} \nonumber \\
=&\; \int_{k}dk \Psi^{\ms{T}}_kH(k) \Psi_{k},
\end{align}
where in the second line, we have introduced the spinor $\Psi^{\ms{T}}=(\chi_1, \bar{\chi}_1, \chi_2, \bar{\chi}_2,\chi_3, \bar{\chi}_3)$ and transformed to momentum space. The explicit form of $H(k)$ is
\begin{align}
H(k)= vk \mathrm{Id}_{3}\otimes \sigma^{z}+ \sum_{j}i m_{j}\mathrm{L}^{j}\otimes \sigma^{y}.
\end{align}
The spectrum of $H(k)$ is gapless with the eigenvectors $|\psi_{1}\rangle =(m_1,m_2,m_3)^{\ms{T}}\otimes (1,0)^{\ms{T}}$ and $|\psi_{2}\rangle =(m_1,m_2,m_3)^{\ms{T}}\otimes (0,1)^{\ms{T}}$ having eigenvalues $\pm vk$. We are able to find the above two vectors due to the simple condition that
\begin{align}
    \text{Ker}\Big[\sum_{j} m_{j}\mathrm{L}^{j}\Big]\neq \varnothing,
\end{align}
which follows from the fact that $\mathrm{L\equiv} \sum_{j}m_{j}\mathrm{L}^{j}$ is a generic $3\times 3$ anti-symmetric matrix, and is therefore singular, since
\begin{equation}
\mathrm{det}(\mathrm{L}^{\ms T})=\mathrm{det}(-\mathrm{L})=(-1)^3\mathrm{det}(\mathrm{L}).
\end{equation}
The above argument can be directly generalized to any odd number of helical modes, since the corresponding $\mathrm{L}$ will always be singular, and result in the existence of gapless eigenvectors. More generally, we may use a mathematical theorem \citep{yangrmp1962} that states any anti-symmetric matrix $\mathrm{L}$ can be block diagonalized by conjugating with an orthogonal matrix. We can then verify that for an even number of helical modes, we can block diagonalize the Hamiltonian into $2\times 2$ blocks with a gapped spectrum.

Having established the stability of an odd number of helical modes at the non-interacting level, we proceed to examine the effect of four fermion or current-current interaction terms for the $\mathfrak{so}(3)_1\times \overline{\mathfrak{so}(3)}_1$ theory. The action of TRS on the currents is 
\begin{align}
    \mathcal T: \mathcal J^{\ms{a}} \longleftrightarrow -\bar{\mathcal J}^{\ms{a}}.
\end{align}
Therefore the general form of TRS invariant current-current interaction terms is
\begin{equation}
\delta {H}_{\ms{int}}=\sum_{\ms{a}}\lambda_{\ms{a}}\mathcal J^{\ms{a}}\bar{\mathcal J}^{\ms{a}}+\sum_{\ms{a,b}}\lambda_{\ms{ab}}\left(\mathcal J^{\ms{a}}\bar{\mathcal J}^{\ms{b}}+\bar{\mathcal J}^{\ms{a}}{\mathcal J}^{\ms{b}}\right).
\end{equation}
We examine the $\lambda_{\ms{a}}$ terms first. The term $\mathcal J^{\ms{a}}\bar{\mathcal J}^{\ms{a}}$ can be decomposed into two kinds of bilinears: those of the form $\chi^{j}\bar{\chi}^{j}$, $j\neq \ms{a}$, and those of the form $\mathcal O^{\ms{a}}$. In other words if $\lambda_{\ms{a}}$ were to flow to strong coupling, at least one of these bilinears would be expected to acquire a groundstate expectation value. Since the former kind breaks TRS, this would lead to a groundstate that spontaneously breaks TRS. Alternatively, we could consider the scenario where $\mathcal O^{\ms{a}}$ acquires an expectation value. An important observation is that
\begin{equation}
\left(\mathcal O^j\right)^2\propto \mathcal J^{j}\bar{\mathcal J}^{j},
\end{equation}
up to a constant term. Therefore, by ramping up $\lambda^{1}$ for example, we can gap out the modes $\chi_{2,3}$ and $\bar{\chi}_{2,3}$ by condensing $\mathcal O^{\ms{1}}\propto (\chi_{2}\bar{\chi_{3}}+\bar{\chi}_{2}\chi_{3})$. Crucially though, we cannot gap out the entire theory by simultaneously condensing $\mathcal O^{1,2,3}$, as these operators satisfy the non-trivial algebra in Eq.~\eqref{eq:non_abelian_haldane}. This can be understood as a generalization of the Haldane criterion to non-Abelian current algebras. 

 Next, we turn to the terms of the form $\lambda_{\ms{ab}}(\mathcal J^{\ms{a}}\bar{\mathcal J}^{\ms{b}}+\bar{\mathcal J}^{\ms{a}}\mathcal J^{\ms{b}})$. We are interested in how the groundstate at $\lambda_{\ms{ab}}\to \infty$ transforms under TRS. To this end, we decouple the interaction term into possible products of fermion bilinears and ask whether we can find a decoupling where each bilinear is invariant under TRS. Let us illustrate this procedure with an explicit example. Consider the term
\begin{align}
\mathcal J^{1}\bar{\mathcal J}^2+\bar{\mathcal J}^1{\mathcal {J}}^2=&\;(i\chi_2\chi_3)(i\bar{\chi}_1\bar{\chi}_3)+(i\bar{\chi}_2\bar{\chi}_3)(i\chi_1\chi_3)\nonumber \\
=&\; (i\chi_{3}\bar{\chi}_{3})(i\bar{\chi}_{1}\chi_{2}-i\chi_{1}\bar{\chi}_2).
\end{align}
This is the only possible decoupling for the term proportional to $\lambda_{12}$; the terms proportional to the other $\lambda_{\ms{ab}}$'s all have similarly unique decouplings. Crucially, both the bilinears in the decoupling transform non-trivially under TRS and such an interaction cannot have a TRS invariant groundstate. 

 The above considerations generalize to any odd number of helical modes. For an even number of helical modes, say $2n$, we can construct an interaction term that gaps out all the modes while preserving the TRS. Consider the matrices $\widetilde{\mathrm{L}}_{2n}^{\ms{a}}$ with $\ms{a}=1,\dots,n$ which generate a $\mathfrak{so}(2)^{n}$ subgroup $\mathfrak{so}(2n)$. The matrix $\mathrm{L}_{2n}^{\ms{a}}$  basically generates rotations in ${x}_{2\ms{a}-1}\text{-}x_{2\ms{a}}$ plane in $\mathbb R^{2n}$. Then we can construct the currents
\begin{align}
    \mathcal J^{\ms{a}}:=\frac{i}{2}\chi_{j}\widetilde{\ms{L}}_{2n,jk}^{\ms{a}}\chi_{k}=i\chi_{2\ms{a}-1}\chi_{2\ms{a}},
\end{align}
and analogously we define the antiholomorphic currents $\bar{\mathcal {J}}^{\ms{a}}$. Then we may write down the interaction term
\begin{align}
    \delta H=&\; \lambda\sum_{\ms{a}=1}^{n}\mathcal J^{\ms{a}}\bar{\mathcal J}^{\ms{a}} \nonumber \\
= &\; \lambda\sum_{\ms{a}=1}^{n} (i\chi_{2\ms{a}-1}\bar{\chi}_{2\ms{a}}+i\bar{\chi}_{2\ms{a}-1}\chi_{2\ms{a}})^{2} + \text{const.}.
\end{align}
Since the terms $(i\chi_{2\ms{a}-1}\bar{\chi}_{2\ms{a}}+i\bar{\chi}_{2\ms{a}-1}\chi_{2\ms{a}})$ are TRS invariant and commute mutually for all $\ms{a}$, adding such a term gaps out all $2n$ helical modes simultaneously.

\subsubsection{Gapping out the surface with topological order} In this subsection, we describe the STO for second-order 3D class DIII HOTSC protected by inversion symmetry. Here we start with the system that originally carries three pairs of helical Majorana modes along the hinge. Since the class DIII hinge modes can be regarded as a stack of class D hinge modes with their time-reversed partners, a natural candidate of the STO for class DIII is given by stacking the STO for class D with its time-reversed partner which is $\mathsf{SO}(3)_3\times \overline{\mathsf{SO}(3)_3}$. 

 We consider the HOTSC to have a spherical geometry with three pairs of helical Majorana modes $\chi_i,\bar{\chi}_i$ propagating along the equator on the surface. First we add three copies of class DIII superconductors on the northern and southern hemisphere, ending up with $6$ additional pairs of helical Majorana modes on the equator which we label as $\gamma^{\mathrm{N/S}}_i,\bar{\gamma}^{\mathrm{N/S}}_i$. The hinge is  described by the $\mathfrak{so}(9)_1\times \overline{\mathfrak{so}(9)}_1$ WZW theory. The action of inversion on the different Majorana operators is as follows
\begin{align}
    \mathcal I:&\;
    \begin{bmatrix}
    \chi_i \\
    \gamma_{i}^{\mathrm{N}} \\
    \gamma_i^{\mathrm{S}} \\
    \bar{\chi}_{i} \\
    \bar{\gamma}^{\mathrm{N}}_{i} \\
    \bar{\gamma}^{\mathrm{S}}_{i}
    \end{bmatrix}(\theta)
    \longrightarrow
    \begin{bmatrix}
    i\chi_i \\
    i\gamma_{i}^{\mathrm{S}} \\
    i\gamma_i^{\mathrm{N}} \\
    -i\bar{\chi}_{i} \\
    -i\bar{\gamma}^{\mathrm{S}}_{i} \\
    -i\bar{\gamma}^{\mathrm{N}}_{i}
    \end{bmatrix}(\theta + \pi).
    \label{eq:inv_action_conformal_embedding}
\end{align}
Next, we carry out the conformal embedding procedure i.e. we embed $\mathfrak{so}(3)_3\times \mathfrak{so}(3)_3 \subset \mathfrak{so}(9)_1$ and $\overline{\mathfrak{so}}(3)_3\times \overline{\mathfrak{so}}(3)_3 \subset \overline{\mathfrak{so}}(9)_1$. Since the recipe is identical to that described for the holomorphic CFT in Sec.~\ref{Sec:ClassD_HOTSC}, we do not repeat the procedure here. Eventually, we end up with chiral and anti-chiral current operators that transform under the inversion-symmetry action as
\begin{align}
    \mathcal I:&\;
    \begin{bmatrix}
    \mathcal J^{\ms{a},1} \\
    \bar{\mathcal J}^{\ms{a},1} 
    \end{bmatrix}(\theta)
    \longmapsto
        \begin{bmatrix}
    \mathcal J^{\ms{a},2} \\
    \bar{\mathcal J}^{\ms{a},2} 
    \end{bmatrix}(\theta+\pi),
\end{align}
where `$\ms{a}$' labels the generators of $\mathfrak{so}(3)$. Similarly, under TRS, the currents transform as
\begin{align}
    \mathcal T: \mathcal J^{\ms{a},\kappa} \longleftrightarrow -\bar{\mathcal J}^{\ms{a},\kappa},
\end{align}
where $\kappa\in \left\{1,2\right\}$. In order to gap out these current operators, we introduce topological order $\mathcal A_{\mathrm{N}}$ and $\mathcal A_{\mathrm{S}}$ on the northern and southern hemispheres, respectively, with $\mathcal A_{\mathrm{N}}=\ms{SO}(3)_3\times \overline{\ms{SO}(3)}_3$. Conveniently, the symmetry transformation properties of the edge modes induced on the hinge from the topological order are identical to those of the aforementioned modes obtained from the conformal embedding procedure. We denote the modes provided by the topological orders on the northern and southern hemispheres with subscripts $\mathrm{N}$ and $\mathrm{S}$. Under inversion and TRS,
\begin{align}
    \mathcal I:&\;\mathcal J^{\ms{a}}_{\mathrm{N,S}}(\theta)\longmapsto \mathcal J^{\ms{a}}_{\mathrm{S,N}}(\theta+\pi), \nonumber \\
   \mathcal I:&\; \bar{\mathcal J}^{\ms{a}}_{\mathrm{N,S}}(\theta)\longmapsto \bar{\mathcal J}^{\ms{a}}_{\mathrm{S,N}}(\theta+\pi), \nonumber \\
    \mathcal T:&\; \mathcal J^{\ms{a}}_{\mathrm{N,S}}(\theta)\longleftrightarrow -\bar{\mathcal J}^{\ms{a}}_{\mathrm{N,S}}(\theta).
\end{align}
The hinge modes $\mathcal{J}^{\ms{a},\kappa}, \bar{\mathcal{J}}^{a,\kappa}$ and the edge modes of the surface topological order $\mathcal{J}^{\mathsf{a}}_{\mathrm{N/S}}, \bar{\mathcal{J}}^{\ms{a}}_{\mathrm{N/S}}$ can together be gapped out upon adding the gapping term
\begin{align}
\delta \mathcal{L}&=\lambda(\theta)\sum_{\ms{a}=1}^{3}\Big[\bar{\mathcal J}^{\ms{a}}_\mathrm{N}(\theta)\mathcal J^{\ms{a},1}(\theta)+\bar{\mathcal J}^{\ms{a}}_\mathrm{S}(\theta)\mathcal J^{\ms{a},2}(\theta)\nonumber \\
&+\mathcal J^{\ms{a}}_\mathrm{N}(\theta)\bar{\mathcal J}^{\ms{a},1}(\theta)+\mathcal J^{\ms{a}}_\mathrm{S}(\theta)\bar{\mathcal J}^{\ms{a},2}(\theta)\Big],
\end{align}
which is TRS invariant, and inversion symmetric if $\lambda(\theta)=\lambda(\theta+\pi)$. Therefore, we can choose $\lambda$ to be constant.

\section{Surface Topological Order for Third Order Topological Phases}\label{Sec:3rdOrder}
In this subsection, we discuss the surface topological order for third order topological phases protected by inversion symmetry in addition to possible Altland-Zirnbauer symmetries. In total there are five Altland-Zirnbauer classes that support non-trivial third-order topology upon imposing inversion symmetry. These are D, BDI, AIII, DIII and CII. In what follows we present the STO for classes D, BDI and AIII together, as these classes and consequently their STOs are closely related. 
\begin{figure}[tbh!]
\includegraphics[width=0.3\textwidth]{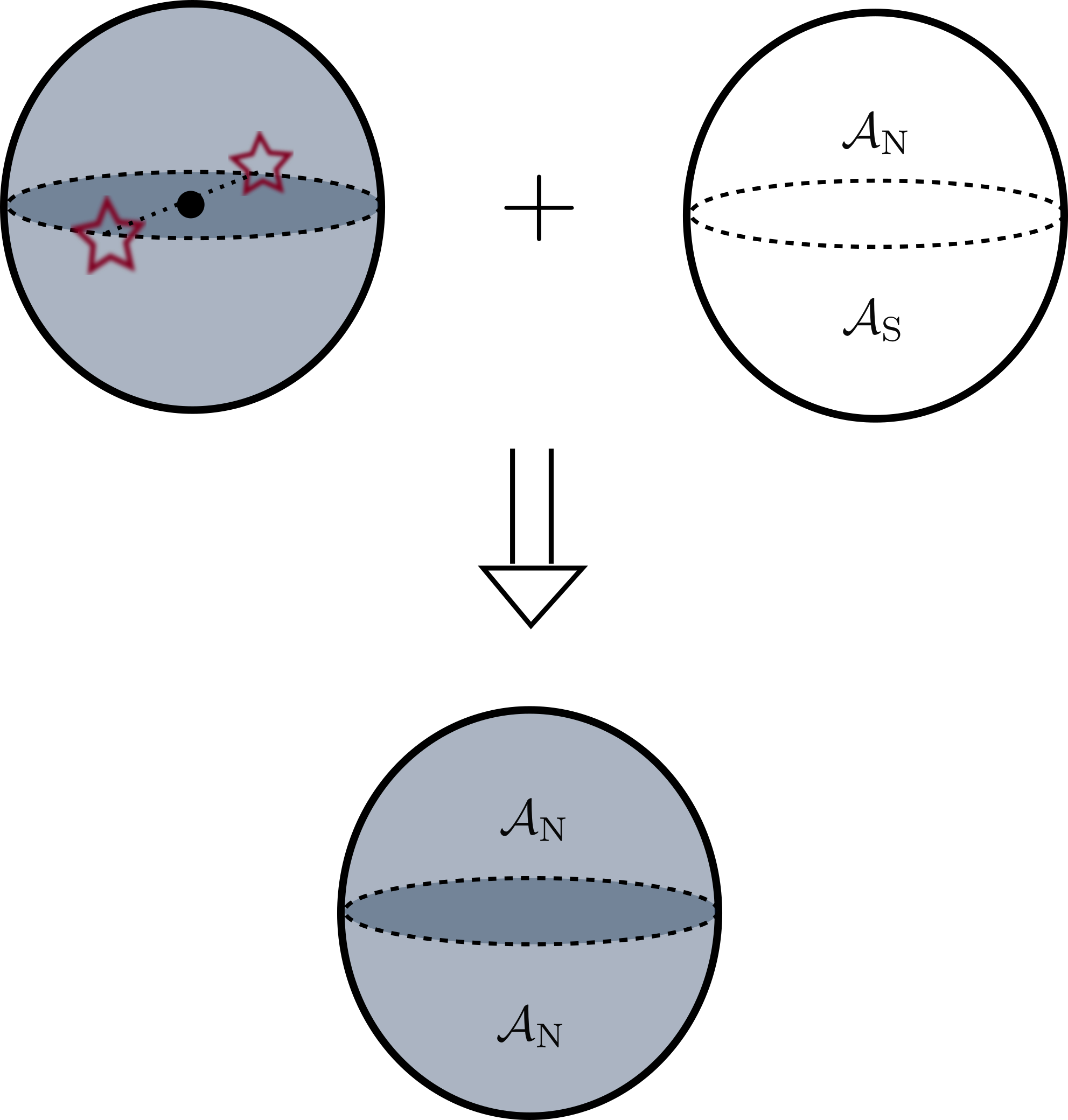}
\caption{Surface topological order for third order topological phases protected by inversion symmetry. In the picture, the red stars denote the zero modes located at antipodal points of the surface.}
\label{STO_3rdOrder_HOTSC}
\end{figure}

\subsection{Class D, BDI and AIII}
We begin with the discussion of class D. Note that the third-order inversion-symmetric class D superconductor can be obtained by stacking two copies of second-order inversion-symmetric class D superconductors whose surface contains the configuration $(1,0)\oplus (1,0)=(2,0)\sim (1,-1)$ in the notation used in Sec.~\ref{Sec:ClassD_HOTSC}. The configuration $(1,-1)$ contains a pair of counter-propagating chiral modes $\chi^{+}$ and $\bar{\chi}^{-}$ which are unstable to a mass term $im(\theta)(\chi^{+}(\theta))^{(\dag)}\bar{\chi}^{-}(\theta)$. The inversion symmetry imposes that $m(\theta+\pi)=-m(\theta)$ and consequently, the mass vanishes at two anti-podal points which contain Majorana zero-modes. Furthermore since we can always gap out Majorana modes in pairs, $(2,0)\oplus (2,0)=(4,0)\sim (0,0)$. This agrees with the result \citep{luka,khalaf_inversion} that 3D class D higher-order topological superconductors enriched by inversion symmetry are classified by $\mathbb Z_{4}$ which is an extension of $\mathbb Z_{2}$ (second-order phases) by $\mathbb Z_{2}$ (third-order phases).\par
Here we describe the procedure to gap out two surface Majorana zero modes (or equivalently the (2,0) configuration) by pasting inversion symmetric surface topological order. First we start with the configuration $(-6,-6)$ which is obtained by pasting 6 inversion symmetric copies of $p\pm ip$ on the surface. Next we paste $\mathfrak{so}(3)_3\times \mathfrak{so}(3)_3$ on the northern and southern hemispheres which effectively provides additional modes corresponding to $(12,6)$. Upon such a surface pasting, we end up with $(-6,-6)\oplus (12,6) = (6,0)\sim (2,0)$. Since we can create two Majorana zero modes on antipodal points on the  surface without manipulating the bulk, we can always absorb the surface modes contributed by a third order class D superconductor.\par
Having obtained the STO for class D, we now proceed to discuss the surface topological order for 3D inversion TSC in BDI class. According to Ref.~\onlinecite{khalaf_inversion}, 3D inversion-symmetric TSC in class BDI has a third order phase but no non-trivial first or second order phases. Conveniently, we make use of the notion of \emph{block state} introduced in Ref.~\onlinecite{Hermele_1, Hermele_2}. Generically, a block state $\state{\Psi}$ has the form $\state{\Psi}=\bigotimes_{b\in B}\state{\psi_{b}}$, where $B$ is a collection of blocks, and block $b$ is a $d_b$-dimensional system embedded in a $d$-dimensional space. In our case, the wavefunction of a non-trivial third order phase is physically equivalent
(up to a inversion-symmetric finite depth unitary circuit) to a block state $\state{\Psi}=\state{\Psi_1}\otimes \state{\Psi_a}$ where $\state{\Psi_1}$ denotes a state in the non-trivial phase of the 1D BDI Majorana chain embedded in the 3D space with inversion symmetry, and $\state{\Psi_a}$ denotes a state, describing the rest of the 3D space, in the trivial phase. The 1D BDI Majorana chain can be described by the following Hamiltonian
\begin{equation}
    \hat{H}=\frac{1}{2}\int dx \hat{\psi}^\dag(x) H(x)\hat{\psi}(x),
\end{equation}
where
\begin{equation}
    H(x)=-i\tau_y\partial_x+m(x)\tau_z, \quad \hat{\psi}^\dag=(a^\dag_x,a_x),
\end{equation}
with $a^\dag_{x}$ being a complex fermion creation operator. The Altland-Zirnbauer symmetries are represented as
\begin{equation}
    \mathcal{T}=\mathcal{K},\quad \mathcal{P}=\tau_x\mathcal{K},\quad \mathcal{I}=\tau_z.
\end{equation}
The mass profile $m(x)$ is positive (negative) inside (outside) the chain. If we denote the two edges of this chain as $x^+$ and $x^-$ separately, then Jackiw-Rebbi procedure shows that these two edges host Majorana zero modes $\gamma_{x^+}$ and $\gamma_{x^-}$, which have the form \cite{ruben_2017}
\begin{align}
    \gamma_{x^+}&=\int dx \frac{1}{\sqrt{2}}(a_x+a^\dag_x)e^{\int^x_{x_a}dx'm(x')},\nonumber \\
    \gamma_{x^-}&=\int dx \frac{i}{\sqrt{2}}(-a_x+a^\dag_x)e^{-\int^x_{x_b}dx'm(x')},
\end{align}
where $x_{a/b}$ are parameters that ensure the normalization condition $\int dx e^{2\int^x_{x_a}dx'm(x')}=\int dx e^{-2\int^x_{x_b}dx'm(x')}=1$.
The actions of TRS and inversion are as follows
\begin{equation}
    \mathcal{T}\gamma_{x^+}\mathcal{T}^{-1}=\gamma_{x^+}, \quad \mathcal{T}\gamma_{x^-}\mathcal{T}^{-1}=-\gamma_{x^-}.\label{eq:trsbdi}
\end{equation}
\begin{equation}
    \mathcal{I}\gamma_{x^+}\mathcal{I}^{-1}=i\gamma_{x^-},\quad  \mathcal{I}\gamma_{x^-}\mathcal{I}^{-1}=-i\gamma_{x^+}.\label{eq:invbdi} 
\end{equation}
The above symmetry actions satisfy the basis invariant relations $\{\mathcal{I}, \mathcal{P}\}=0$, $\left[\mathcal I,\mathcal T\right]=0$, $\mathcal T^{2}=1$ and $\mathcal I^{2}=1$, independently agreeing with Ref.~\onlinecite{khalaf_inversion}. We thus propose that the STO for 3D third order inversion TSC in BDI class to be $\mathsf{SO}(3)_3\times\overline{\mathsf{SO}(3)_3}$, which is the same as we found for the 3D third order inversion TSC in D class. Recall that in Section \ref{Sec:ClassD_HOTSC}, we showed that an inversion symmetric surface realization of $\mathsf{SO}(3)_3$ contains a gapless modes (denoted as $\mathcal{J}^{\ms{a}}_{\mathrm{N/S}}$ currents) appearing on an inversion-symmetric line (chosen as the equator for our convenience). We then used such a current mode to gap out the 9 chiral Majorana modes on the equator contributed by the bulk as well as the surface pasting of $p\pm ip$ superconductors. 
Similarly, $\mathsf{SO}(3)_3\times\overline{\mathsf{SO}(3)_3}$, which is the proposed STO for class BDI, has 9 pairs of counter-propagating Majorana modes, without the contribution from the bulk. Let us denote these modes as $\alpha_i,\bar{\alpha}_i$ with $i=1,\dots,9$. The TRS action on the Majorana modes is
\begin{equation}
    \mathcal{T}: i\mapsto -i, \begin{bmatrix}
    \alpha_i\\
    \bar{\alpha}_i
    \end{bmatrix}\mapsto \begin{bmatrix}
    \bar{\alpha}_i\\
    \alpha_i
    \end{bmatrix}.
\end{equation}
While the inversion acts naturally on the current operators contributed by the STO on the equatorial hinge as 
\begin{align}
\mathcal{I}:
\begin{bmatrix} 
\mathcal{J}^{\ms{a}}_\mathrm{N} \\
\bar{\mathcal{J}}^{\ms{a}}_\mathrm{N}
\end{bmatrix} (\theta)
\longleftrightarrow 
\begin{bmatrix}
\mathcal{J}^{\ms{a}}_\mathrm{S} \\
\bar{\mathcal{J}}^{\ms{a}}_\mathrm{S}
\end{bmatrix}(\theta+\pi).
\end{align} 
After the conformal embedding, we regroup the Majoranas as $\chi_i,\gamma^{(1)}_i,\gamma^{(2)}_i,\bar{\chi}_i,\bar{\gamma}^{(1)}_i,\bar{\gamma}^{(2)}_i$, where the definition is taken as
\begin{align}
\chi_{1,2,3}\equiv&
\alpha_{1,2,3}, \nonumber \\ 
\gamma^{(1)}_{1,2,3}\equiv&\; 
\alpha_{4,5,6}, \nonumber \\
\gamma^{(2)}_{1,2,3}\equiv&\; \alpha_{7,8,9},
\end{align}
and similarly for the $\bar{\chi}_i,\bar{\gamma}^{(1)}_i,\bar{\gamma}^{(2)}_i$. Inversion action on the Majoranas is
\begin{align}
    \mathcal I:&\;
    \begin{bmatrix}
    \chi_i \\
    \gamma_{i}^{(1)} \\
    \gamma_i^{(2)} \\
    \bar{\chi}_{i} \\
    \bar{\gamma}^{(1)}_{i} \\
    \bar{\gamma}^{(2)}_{i}
    \end{bmatrix}(\theta)
    \longrightarrow
    \begin{bmatrix}
    i\chi_i \\
    i\gamma_{i}^{(2)} \\
    i\gamma_i^{(1)} \\
    -i\bar{\chi}_{i} \\
    -i\bar{\gamma}^{(2)}_{i} \\
    -i\bar{\gamma}^{(1)}_{i}
    \end{bmatrix}(\theta + \pi).
\end{align}
We now proceed to gap these modes in groups. For the $\gamma$'s and $\bar{\gamma}$'s, we can write the following gapping term
\begin{equation}
    \delta \hat{H}_1=\sum_i m_1(\theta)(i(\gamma_i^{(1)})^{(\dag)}\bar{\gamma}_i^{(1)}-i(\gamma_i^{(2)})^{(\dag)}\bar{\gamma}_i^{(2)}),
\end{equation}
which gaps all the $\gamma$ and $\bar{\gamma}$. For the $\chi$'s, $\bar{\chi}$'s, we can write the following gapping term:
\begin{equation}
    \delta \hat{H}_2=\sum_im_{2,i}(\theta)(i\chi^{(\dag)}_i\bar{\chi}_i).
\end{equation}
Note that inversion forces $m_{2,i}(\theta)=-m_{2,i}(\theta+\pi)$, and each counter propagating $\chi,\bar{\chi}$ contributes a pair of Majorana zero modes. Since we have three pairs of counter-propagating $\chi,\bar{\chi}$, two pairs of Majorana zero modes will be gapped out, and we are therefore left with one protected pair of Majorana zero modes on the equator located at inversion-symmetric positions. Furthermore, the Jackiw-Rebbi procedure shows that these Majorana zero modes transform in the exact same way as in Eq.~\eqref{eq:trsbdi} and Eq.~\eqref{eq:invbdi}. Thus the zero modes from the STO can gap out the zero modes from the BDI bulk. 

What about the third order inversion symmetric topological phase in AIII? According to Ref.~\onlinecite{khalaf_inversion}, a 3D inversion symmetric TI in class AIII has only a third-order implementation. Physically, this phase can be thought of as a 1D AIII Su-Schrieffer-Heeger (SSH) chain which is in non-trivial phase inserted into a 3D manifold. There is a close connection between the 1D SSH chain and the 1D BDI chain, as pointed out in Ref.~\onlinecite{ruben_2017}. we can establish an exact mapping from two copies of the BDI Kitaev chain to one copy of SSH chain. The two dangling Majorana zero modes form a Dirac zero mode, which is the dangling zero mode of the SSH chain. Since two copies of Kitaev chain have an emergent $\ms{O}(2)$ symmetry, its subgroup $\ms{SO}(2)$ corresponds to the $\ms{U}(1)$ symmetry for the AIII chain. Crucially, the TRS in the BDI chain corresponds to the sub-lattice/chiral symmetry of the AIII chain (We comment that the TRS is still anti-unitary in the Fock space after the mapping, but it is unitary on the single-particle Hamiltonian \cite{ruben_2017}). Therefore, we naturally conclude that the STO for the 3D third order AIII phase is equivalent to the STO for two copies of 3D third order BDI phase, with chiral symmetry implemented in the same way as the TRS in the STO for BDI phase. 
\subsection{Class DIII}
In this subsection, we discuss the STO for third order class DIII HOTSC. To that end, we first demonstrate the fact that, for 3D class DIII HOTSC protected by inversion, the classification is $\mathbb{Z}_4$, which is an extension of $\mathbb{Z}_2$ (second-order phases) by $\mathbb{Z}_2$ (third-order phases). Consider the second-order case, where there is a pair of helical hinge modes. We denote the Majorana hinge modes as $(\chi,\bar{\chi})$. The TRS action on these modes is
\begin{align}
    \mathcal T:
    \begin{bmatrix}
    \chi(\theta)\\
    \bar{\chi}(\theta)
    \end{bmatrix}\mapsto 
    \begin{bmatrix}
    \bar{\chi}(\theta)\\
    -\chi(\theta)
    \end{bmatrix},
    \label{eq:TRS_Majorana}
\end{align}
while the inversion action on these modes is
\begin{align}
    \mathcal I:
    \begin{bmatrix}
    \chi(\theta)\\
    \bar{\chi}(\theta)\\
    \chi(\theta+\pi)\\
    \bar{\chi}(\theta+\pi)
    \end{bmatrix}\mapsto 
    \begin{bmatrix}
    i\chi(\theta+\pi)\\
    -i\bar{\chi}(\theta+\pi)\\
    -i\chi(\theta)\\
    i\bar{\chi}(\theta)
    \end{bmatrix}.
    \label{eq:inv_repn_maj}
\end{align}
As in the case of Class D, the above action can be derived using recursive Jackiw-Rebbi procedures. Now suppose we have two copies of such helical modes $(\chi_1,\bar{\chi}_1,\chi_2,\bar{\chi}_2)$, with symmetry action exactly the same as the above. A gapping term can be written down in the 1D model
\begin{equation}
    \delta \hat{h}=\int \mathrm{d}\theta \left[im(\theta)\left(\chi^{(\dag)}_1(\theta)\bar{\chi}_2(\theta)+\bar{\chi}^{(\dag)}_1(\theta)\chi_2(\theta)\right)\right].
\end{equation}
The above term is TRS invariant, and inversion symmetry imposes $m(\theta)=-m(\theta+\pi)$. Thus the 1D system is gapped out except at two inversion symmetric points. Furthermore, these two point modes can be gapped out if we take a double stacking of this model. We therefore conclude that the class DIII HOTSC protected by inversion has classification $\mathbb{Z}_4$, which is an extension of $\mathbb{Z}_2$ (second-order phases) by $\mathbb{Z}_2$ (third-order phases).\par
To gap out the third order topology, we make use of the fact that the third order phase is obtained by two copies of second order phase. Now since the STO for the second order phase is $\mathsf{SO}(3)_3\times\overline{\mathsf{SO}(3)_3}$, we conclude that the STO for third order topological phase is two copies of $\mathsf{SO}(3)_3\times\overline{\mathsf{SO}(3)_3}$ topological order.
\subsection{Class CII}
We now briefly discuss the STO for the third order 3D HOTSC in class CII. Similarly to the case AIII, we can view the 3D inversion-symmetric third order CII phase as a 1D inversion-symmetric CII chain embedded in a 3D manifold. The 1D CII chain always has even number of Majorana zero modes at its edge, instead of single Majorana zero mode at the edge of Kitaev chain. we can view the edge zero modes for CII chain as a Kramers' pair of Majorana zero modes, similar to the edge mode in the case of 1D DIII chain. However 1D CII chain has a $2\mathbb{Z}$ classification whereas the 1D DIII chain has a $\mathbb{Z}_2$ classification \cite{RMP_ChingKai}. Here we briefly look at the following linearised 1D model for CII chain taken from Ref.~\onlinecite{cii_wang_zhao}
\begin{equation}
H(k)=-k\tau_z\sigma_y+m\tau_x, \quad \mathcal{T}=\sigma_y \mathcal{K}, \quad \mathcal{P}=\tau_y \mathcal{K}.
\end{equation}
The chain is in the topological phase when $m>0$,  and the zero modes are trapped at the $m=0$ domain wall. We perform the Jackiw-Rebbi projection to track the symmetry action on the zero modes, and found that $\mathcal{T}_{\mathrm{edge}}\sim \mathcal{P}_{\mathrm{edge}}\sim i\sigma_y \mathcal{K}$. Because of this, stacking any number of copies of CII chains cannot enable us to gap these zero modes out, as TRS must commute with the mass term, whereas the PHS must anti-commute with the mass term. The fact that CII can be viewed as a stacking of a Kitaev chain and its TRS copy with $\mathcal{P}^2=-1$ implementation of PHS leads us to conjecture that the STO for the third order 3D inversion CII phase to be the same as the STO for the third order 3D DIII phase, with the $\mathcal{P}^2=-1$ implementation of PHS on the STO level.

\section{Summary and outlook}

In this work, we have established via an explicit construction that all inversion-symmetric higher-order topological insulators and superconductors (except for AZ class C) admit gapped surfaces with anomalous topological order.
While we have done so for the case of inversion symmetric electronic phases, we expect it to hold more generally for both Bosonic and electronic three-dimensional higher-order phases with various spatial symmetries. 
This consequently extends the list of symmetric surface terminations of 3D second and third-order topological phases to include `anomalous gapped surfaces'. 
A technical consequence of this work is the study of spatial symmetries and their anomalies in 2D topological orders. 
In this regard, there is a recent systematic algebraic framework to study the spatial symmetry enrichment of modular tensor categories \cite{maissam_new_4}. 
In particular, obstructions to such symmetry enrichment must precisely encode anomalies that in turn can be compensated by crystalline topological phases in one higher dimension. 
It would be interesting to pursue this direction in future work.

\section*{Acknowledgements}
We would like to thank Eslam Khalaf for useful comments. 
We acknowledge support from the European Research Council under the European Union Horizon 2020 Research and Innovation Programme, through Starting Grant [Agreement No. 804213-TMCS, ML and SP] and [ERC-StG-Neupert757867-PARATOP, TN], the Marie Sklodowska
Curie Action [Grant Agreement No 701647, AT], the Swedish Research Council (VR) through grants number 2019-04736 and 2020-00214
and from a Buckee Scholarship at Merton College [ML]. 
Finally, we acknowledge Alexander, Serena, and Amaia, without whose arrival into our lives  this paper would have been finished in half the time but with a tenth the joy.

\appendix

 \section{$K$-matrix Luttinger liquids}\label{appendix:k-matrix}
 In this appendix, we briefly review the $K$-matrix theory of Luttinger liquids \cite{Wen_1990, Wen_1995, wen2004quantum}. 2D Abelian topological orders can be described by $n$ emergent $\mathsf{U}(1)$ gauge fields coupled via a Chern-Simons action \citep{xiao-gang_zee}. If the 2D system has a boundary, the boundary can be described by $\mathsf{U}(1)$ compact boson theory \citep{Wen_1995}. Specifically, the bulk Lagrangian and the boundary Lagrangian can be written as:
 \begin{align}
 \mathcal{L}_{\text{bulk}}&=\frac{K_{IJ}}{4\pi}\epsilon_{\mu\nu\sigma}a^{I,\mu} \partial^\nu a^{J,\sigma},\nonumber \\
 \mathcal{L}_{\text{boundary}}&=K_{IJ}\partial_t\phi^I\partial_x\phi^J-V_{IJ}\partial_x\phi^I\partial_x\phi^J\nonumber\\
 &+\sum_j g_j\cos [l^{\mathsf{T}}_{j,I}\phi^I+\alpha].
 \end{align}
 In the bulk Lagrangian, the $K_{IJ}$ is a symmetric integer matrix which describes the Chern-Simons coupling of emergent gauge fields $a^I_{\mu}$. By requiring the gauge invariance of the theory on a $2+1d$ manifold with boundary, the boundary must carry degrees of freedom $\phi^I$ described by the corresponding boundary action. The sine-Gordon terms are derived by local Hermitian gapping terms i.e. $\cos [l^{\mathsf{T}}_{j,I}\phi^I+\alpha]\sim e^{il^{\mathsf{T}}_{j,I}\phi^I}+e^{-il^{\mathsf{T}}_{j,I}\phi^I}$, where $l^{\mathsf{T}}$s are integer vectors (usually referred to as the gapping vectors). For the convenience of discussion, we also introduce the concept boundary gapping lattice $\Gamma^\partial=\{ l_j\}$, namely the lattice spanned by gapping vectors.\par
 The advantage of focusing on the $K$-matrix is that we can represent quasiparticles in a convenient algebraic way. Let us denote the order of the $K$-matrix as $n$, i.e., there are $n$ gauge fields in the bulk, and a quasiparticle can be represented by a $n$-components vector $l$. The braiding phase between two quasiparticles is given by $\theta_{ll'}=2\pi lK^{-1}l'$, and the topological spin (exchange phase) of the quasiparticle $l$ is given by $\theta_l=\theta_{ll}/2=\pi lK^{-1}l$. To identify local particles i.e. bosons/fermions in the theory, we require the local particle to braid trivially with all particles, thus resulting in the constraint $l=K\Lambda$, where $\Lambda$ is an integer vector.\par
 To have a fully gapped boundary, there are certain criteria that the sine-Gordon terms have to satisfy. More concretely, we look at:
 \begin{equation}
 \delta \mathcal{L}_{\text{boundary}}=\sum_j g_j\cos [l^{\mathsf{T}}_{j,I}\phi^I+\alpha].
 \end{equation}
 Physically, by writing such terms, quasiparticles $l_{j}$s are condensed on the boundary. For all quasiparticles to condense, we require the following conditions \citep{xiao-gang_juven}
 \begin{enumerate}
     \item The condensed quasiparticles have bosonic self-statistics: $\forall l_j\in \Gamma^\partial, l^{\mathsf{T}}_{j,I}K^{-1}_{IJ}l_{j,J}\in 2\mathbb{Z}$.
     \item The condensed quasiparticles mutually braid trivially.
     \item The bosonic fields corresponding to the condensed quasiparticles can acquire classical values at the same time: $\forall l_j,l_i\in \Gamma^\partial, l^{\mathsf{T}}_{i,I}K^{-1}_{IJ}l_{j,J}=0$. This condition is also known as the \emph{Haldane criterion}.
     \item The condensed quasiparticles must be local/non-fractional particles: $\forall l_{j,I}, l_{j,I}=K_{IJ}\Lambda_{j,J}$, where $\Lambda_{j,J}$ is an integer vector.
     \item Completeness: $\forall l_{j,I}=K_{IJ}\Lambda_{j,J}$, if $l^{\mathsf{T}}_{j,I}K^{-1}_{IJ}l_{j,J}=0$, and $l^{\mathsf{T}}_{j,I}K^{-1}_{IJ}l_{i,J}=0$ for $\forall l_{i,J}\in \Gamma^\partial$, then $l_{j,I}\in \Gamma^\partial$.
     \item Non-chirality: the boundary theory must have $p$ left movers and $p$ right movers to begin with.
 \end{enumerate}
 So far we have ignored the existence of global symmetries. It is beyond the scope of this appendix to introduce a complete inclusion of symmetry in the $K$-matrix formalism. We wish only to describe the more relevant symmetry to this paper here. Crucially, we usually have a global $\mathsf{U}(1)$ symmetry if the system under discussion is a fermionic insulating system, e.g., quantum Hall systems and topological insulators. The presence of the global $\mathsf{U}(1)$ symmetry is usually signalled by the coupling of the original degrees of freedom to a background gauge field $A^I$:
 \begin{align}
 \delta \mathcal{L}_{\text{bulk},\mathsf{U}(1)}&=-\frac{1}{2\pi}t_I\epsilon_{\mu\nu\sigma}A^{\mu} \partial^\nu a^{I,\sigma} \nonumber \\
 \delta \mathcal{L}_{\text{boundary},\mathsf{U}(1)}&=\frac{1}{2\pi}t_I\epsilon_{\mu\nu}\partial^\mu \phi^I A^\nu,
 \end{align}
 where $t_I$ is an integer vector usually referred to as the electric charge vector. Quasiparticle $l$'s electric charge is given by $q_l=\frac{1}{2\pi}l^{\mathsf{T}}K^{-1}t$. Upon the introduction of the global $\mathsf{U}(1)$ symmetry, the quasiparticle condensation on the boundary is further restricted: only the charge neutral particles can be condensed ,i.e., $\forall l_j, l_j K^{-1}t=0$.

 \section{$\mathcal T$-Pfaffian topological order}
 In this appendix, we briefly review the $\mathcal T$-Pfaffian topological order proposed in Ref.~\onlinecite{Bonderson_2013, Chen_2014,Wang_Senthil_2014}. The $\mathcal T$-Pfaffian topological order was proposed as a possible symmetric interacting surface phase of the 3D topological insulator. Similar to a 3DTI, the $\mathcal T$-Pfaffian is $\mathbb{Z}_2$  classified in the sense that two copies of $\mathcal T$-Pfaffian is can be condensed into a trivial state. The $\mathcal T$-Pfaffian topological order can be viewed as a product of two topological order:
 \begin{equation}
 \text{T-Pf}\equiv \text{Ising}\times \mathsf{U}(1)_{-8}/\mathbb{Z}_2.
 \end{equation}
 The $\text{Ising}$ topological order has three anyons
 \begin{equation}
 \overline{\text{Ising}}=\{1,\psi,\sigma \},
 \end{equation}
 with fusion rules:
 \begin{equation}
 \sigma\times \psi=\sigma,\sigma\times \sigma=1+\psi,\psi\times \psi=1,
 \end{equation}
 and topological spins:
 \begin{equation}
 \theta_{1}=1,\theta_{\psi}=-1,\theta_{\sigma}=e^{i\frac{\pi}{8}}.
 \end{equation}
 The $\mathsf{U}(1)_{-8}$ topological order has 8 anyons:
 \begin{equation}
 \mathsf{U}(1)_{-8}=\{0,1,2,3,\dots, 7 \},
 \end{equation}
 with fusion rules:
 \begin{equation}
 p\times q=(p+q) \text{ mod } 8,
 \end{equation}
 and topological spins:
 \begin{equation}
 \theta_k=e^{-i\frac{\pi}{8}k^2}.
 \end{equation}
 The product is taken such that $1,\psi\in \text{Ising}$ are combined with even $p\in \mathsf{U}(1)_{-8}$, and $\sigma\in \text{Ising}$ is combined with odd $p\in \mathsf{U}(1)_{-8}$. Therefore we arrive at:
 \begin{equation}
 \text{T-Pf}=\{1_0, 1_2,1_4,1_6, \psi_0, \psi_2,\psi_4,\psi_6,\sigma_1, \sigma_3,\sigma_5,\sigma_7 \}.
 \end{equation}
 The fusion rules and topological spins can be obtained as the product of the $\text{Ising}$ topological order and the $\mathsf{U}(1)_{-8}$ topological order. We list the topological spin in Table.~\ref{table:2}.\par
 \begin{table}[tbh!]
 \begin{center}
 \begin{tabular}{|c|c|c|c|c|c|c|c|c|c|c|c|c|}
 \hline
  & $1_0$& $1_2$& $1_4$& $1_6$& $\psi_0$& $\psi_2$& $\psi_4$& $\psi_6$& $\sigma_1$& $\sigma_3$& $\sigma_5$& $\sigma_7$\\
  \hline
 $\theta$  & $1$& $i$ & $1$& $i$ & $-1$& $-i$ & $-1$& $-i$ & $1$ & $-1$ & $-1$ & $1$\\
 \hline
 \end{tabular}
 \end{center}
 \caption{Topological spins for $\mathcal{T}$-Pfaffian topological order.}\label{table:2}
 \end{table}
 Among the above anyons, $\psi_4$ is special since it is a local object, i.e., it braids trivially with all the other anyons in the theory, has topological spin $-1$ and $\mathsf{U}(1)$ charge $e$. Therefore it is identified as the physical electron. The existence of such a local object is a feature of fermionic topological order, indicating the non-modularity of the theory. The $\mathcal{T}$-Pfaffian is TRS invariant. Most anyons have real topological spins and are invariant under the action of TRS which complex conjugates the spin. The remaining anyons i.e. $1_2,1_6,\psi_2,\psi_6$ transform under TRS as
 \begin{equation}
 \mathcal{T}: 1_2\leftrightarrow \psi_2, 1_6\leftrightarrow \psi_6.
 \end{equation}
 The assignment of $\mathcal{T}^2$ signs is crucial as it is related to the anomaly of the theory. Starting from the requirement $\mathcal{T}^2=-1$ for $\psi_4$, we can obtain two consistent assignments of $\mathcal{T}^2$ signs which are collected in Table.~\ref{table:3}, and the $\mathcal{T}$-Pfaffian with these two assignments are coined $\mathcal{T}$-Pfaffian$_+$ and $\mathcal{T}$-Pfaffian$_-$ respectively.
 \begin{table}[tbh!]
 \begin{center}
 \begin{tabular}{|c|c|c|c|c|c|c|c|c|c|c|c|c|}
 \hline
  & $1_0$& $1_2$& $1_4$& $1_6$& $\psi_0$& $\psi_2$& $\psi_4$& $\psi_6$& $\sigma_1$& $\sigma_3$& $\sigma_5$& $\sigma_7$\\
  \hline
 $\eta$  & $1$& & $-1$& & $1$& & $-1$& & $\pm$ & $\mp$ & $\mp$ & $\pm$\\
 \hline
 \end{tabular}
 \end{center}
 \caption{Symmetry fractionalization pattern for $\mathcal{T}$-Pfaffian$_{\pm}$.}\label{table:3}
 \end{table}
 Using anomaly indicators, authors in  Ref.~\onlinecite{Wang_2017} showed that $\mathcal{T}$-Pfaffian$_+$ is TRS anomaly free while $\mathcal{T}$-Pfaffian$_-$ is TRS anomalous. It is known that, for fermionic SPT protected by $\mathsf{U}(1)\rtimes \mathbb{Z}^{\mathcal{T}}_2$, upon breaking $\mathsf{U}(1)$, the surface becomes trivially gapped. Therefore we do not want the correct STO to possess a TRS anomaly. Thus $\mathcal{T}$-Pfaffian$_+$ is the correct STO for 3D TI.
 By breaking TRS, we can obtain a purely 2D chiral topological order with the same anyon contents as the $\mathcal{T}$-Pfaffian \citep{Bonderson_2013}. The edge of such a 2D chiral topological order contains a Dirac mode and a counter propagating Majorana mode which is described by the following Lagrangian
 \begin{equation}
 \mathcal{L}=\frac{2}{4\pi}\partial_x \phi(\partial_t-v_1\partial_x)\phi+i \gamma(\partial_t+v_2\partial_x)\gamma,
 \end{equation}
where $\phi$ is a chiral compact Boson and $\gamma$ is a Majorana-Weyl mode.

 \section{$\ms{SO}(N)_1$ Wess-Zumino-Witten theory} \label{appendix:wzw}
 In this appendix, we briefly review the chiral $\ms{SO}(N)_1$ WZW conformal field theory which describes $N$ chiral Majorana fermions on a $1+1 d$ manifold \citep{cft_fran}. The $\ms{SO}(N)$ global symmetry arises due to the flavor symmetry of the $N$ chiral Majorana fermions $\chi_i\mapsto O_{ij}\chi_j, O_{ij}\in \ms{SO}(N)$. Since this is a continuous symmetry, there exists corresponding Noether currents:
 \begin{equation}
 J^{\ms{a}}=\frac{i}{2}\chi_l L^{\ms{a}}_{lm}\chi_m, \ms{a}=1,\dots,\frac{N(N-1)}{2},
 \end{equation}
 where $L^{\ms{a}}$s are anti-symmetric $N\times N$ matrices that generate the $\mathfrak{so}(N)$ Lie algebra. \
 Majorana fermions operators have the following OPE:
 \begin{equation}
 \chi_{\ms{a}}(z)\chi_{\ms{b}}(w)=\frac{\delta^{\ms{ab}}}{z-w}+\dots,
 \end{equation}
 from which the OPE for currents is derived:
 \begin{equation}
 J^{\ms{a}}(z)J^{\ms{b}}(w)= \frac{\delta^{\ms{ab}}}{(z-w)^2}+\frac{if_{\ms{abc}}J^{\ms c}(w)}{z-w}+\dots,
 \end{equation}
 where $f_{\ms{abc}}$ is the structure constant for $\mathfrak{so}(N)$. The energy momentum tensor is obtained via the Sugawara construction, and is equivalent to the free fermion energy momentum tensor
 \begin{equation}
 T(z)=\frac{1}{2(N-1)}\vec{J}(z)\cdot \vec{J}(z)=-\frac{1}{2}\sum_{i}\chi_i\partial\chi_i(z).
 \end{equation}
 The OPE of the energy momentum tensor is given by:
 \begin{equation}
 T(z)T(w)= \frac{N/4}{(z-w)^4}+\frac{2T(w)}{(z-w)^2}+\frac{\partial_w T(w)}{z-w}+\dots,
 \end{equation}
 from which we can read off the chiral central charge $c_-=N/2$. A procedure termed the conformal embedding allows us to decompose the original WZW theory into two smaller theories i.e. $\mathfrak{so}(N^2)_1\supseteq \mathfrak{so}(N)^{(1)}_N\times \mathfrak{so}(N)^{(2)}_N$. For illustration purpose, we will review the conformal embedding for $\mathfrak{so}(9)_1\supseteq\mathfrak{so}(3)^{(1)}_3\times \mathfrak{so}(3)^{(2)}_3$.  \
 The original $\mathfrak{so}(9)_1$ theory has $9$ Majorana fermions, denoted by a pair of indices $(i,j),i,j=1,2,3$. We introduce a spinor $\Psi$ to simplify the notation, thus all Majoranas are denoted by $\Psi_{(i,j)}$. The current of the $\mathfrak{so}(9)_1$ theory is given by:
 \begin{equation}
 J^j=\frac{i}{2}\Psi_{(a_1,a_2)}\Sigma^j_{(a_1,a_2),(b_1,b_2)}\Psi_{(b_1,b_2)}, j=1,2,\dots,36,
 \end{equation}
 where $\Sigma^j$ is the generator of the Lie algebra $\mathfrak{so}(9)$. To perform conformal embedding, we consider the following currents:
 \begin{equation}
 J^{j,(\kappa)}=\frac{i}{2}\Psi_{(a_1,a_2)}\sigma^{j,(\kappa)}_{(a_1,a_2),(b_1,b_2)}\Psi_{(b_1,b_2)},
 \end{equation}
 where $ j=1,2,3; \kappa=1,2$
 \begin{align}
 \sigma^{j,(1)}_{(a_1,a_2),(b_1,b_2)}=&\;\Lambda^j_{a_1,b_1}\delta_{a_2,b_2}, \nonumber \\ 
 \sigma^{j,(2)}_{(a_1,a_2),(b_1,b_2)}=&\;\delta_{a_1,b_1}\Lambda^j_{a_2,b_2}.
 \end{align}
 Note that $\Lambda^j$ are the generators for $\mathfrak{so}(3)$. In the fundamental representation, these take the following form
 \begin{equation}
 \Lambda^1=\begin{bmatrix}
 0 & 0 & 0\\
 0 & 0 & 1\\
 0 & -1 & 0
 \end{bmatrix},\Lambda^2=\begin{bmatrix}
 0 & 0 & 1\\
 0 & 0 & 0\\
 -1 & 0 & 0
 \end{bmatrix},\Lambda^3=\begin{bmatrix}
 0 & 1 & 0\\
 -1 & 0 & 0\\
 0 & 0 & 0
 \end{bmatrix}.
 \end{equation}
 Explicitly, the currents of the sub-theories in terms of Majorana fermions are
 \begin{align}
 J^{1,(1)}&=i(\chi_2\chi_3+\chi_5\chi_6+\chi_8\chi_9)\nonumber \\
 J^{2,(1)}&=i(\chi_1\chi_3+\chi_4\chi_6+\chi_7\chi_9)\nonumber \\
 J^{3,(1)}&=i(\chi_1\chi_2+\chi_4\chi_5+\chi_7\chi_8)\nonumber \\
 J^{1,(2)}&=i(\chi_4\chi_7+\chi_5\chi_8+\chi_6\chi_9)\nonumber \\
 J^{2,(2)}&=i(\chi_1\chi_7+\chi_2\chi_8+\chi_3\chi_9)\nonumber \\
 J^{3,(2)}&=i(\chi_1\chi_4+\chi_2\chi_5+\chi_3\chi_6),
 \end{align}
 where we have used the notation
 \begin{align}
 &\Psi_{(1,1)}=\chi_1, \Psi_{(2,2)}=\chi_5, \Psi_{(3,3)}=\chi_9;\nonumber \\
 &\Psi_{(1,2)}=\chi_4, \Psi_{(2,3)}=\chi_8, \Psi_{(3,1)}=\chi_3;\nonumber \\
 &\Psi_{(2,1)}=\chi_2, \Psi_{(3,2)}=\chi_6, \Psi_{(1,3)}=\chi_7.
 \end{align}
 The OPEs computed for these currents takes the form
 \begin{equation}
 J^{\ms{a},(u)}(z)J^{\ms{b},(u)}(w)= \frac{3\delta^{\ms{ab}}}{(z-w)^2}+\frac{i\epsilon_{\ms{ab}j}J^j(w)}{z-w}+\dots,
 \end{equation}
 where the level $3$ is determined by double contraction. Note that $J^{\ms{a},(1)}(z)J^{\ms{b},(2)}(w)$ is non-singular, so that the two sub-theories are decoupled. Thus we conclude that $J^{i,(u)}$ forms $\mathfrak{so}(3)^{(u)}_3$ current algebra. Furthermore, it was shown in Ref.~\onlinecite{teo_majorana} that
 \begin{equation}
 T_{\mathfrak{so}(9)_1}=T_{\mathfrak{so}(3)^{(1)}_3}+T_{\mathfrak{so}(3)^{(2)}_3}.
 \end{equation}
Thus the conformal embedding is complete. 

 \section{$\ms{SO}(3)_3$ topological order}\label{appendix:so anyon model}
 In this appendix, we briefly review the $\ms{SO}(3)_3$ anyon model \citep{Meng_2018}, which is the proposed STO for inversion-symmetric topological superconductors. \par
 The $\mathsf{SO}(3)_3$ anyon model contains anyons $\{0,\frac{1}{2},1,\frac{3}{2},2,\frac{5}{2},3\}$. The fusion rule is given by:
 \begin{equation}
     i\times j=\sum^{\min[i+j,6-(i+j)]}_{k=|i-j|}k.
 \end{equation}
 The topological spin is encoded in the T-matrix, which is given by:
 \begin{equation}
     T=\mathrm{diag}\{1,e^{i\frac{3\pi}{16}},i,e^{i\frac{15\pi}{16}},-i,e^{i\frac{3\pi}{16}},-1 \}.
 \end{equation}
 The quantum dimension of the anyons is listed as the following:
 \begin{equation}
     \{d_i \}=\{1,\sqrt{2+\sqrt{2}},1+\sqrt{2},\sqrt{4+2\sqrt{2}},1+\sqrt{2},\sqrt{2+\sqrt{2}},1 \}.
 \end{equation}
From the above data, we can derive the S-matrix information via the Kitaev ribbon formula. Furthermore, the chiral central charge of this topological order is given by $c=\frac{9}{4}$.
 
 \section{The Jackiw-Rebbi Projection Procedures}
 \label{App:Jackiw_Rebbi}
In a seminal work, Jackiw and Rebbi \cite{jackiw_rebbi} identified a generic mechanism where fermionic zero modes appear localized on the mass domain walls of a 1D Dirac fermionic system. Their approach has been generalized in the condensed matter literature to study zero modes localized at the boundaries of bulk topological phases, e.g., Dirac mode at the end of the SSH chain. In this paper, we have used this procedure to track the symmetry action on the gapless modes of the higher order topological phase. For the sake of brevity, however, we chose not to explicitly show the calculation for every case in this paper, but refer to a more systematic future work \cite{minghao_notes}. Instead, we demonstrate the procedure in this appendix for the second order inversion HOTSC in class D.\par

First, we start with one copy of 3D bulk Hamiltonian of class DIII TSC with inversion symmetry. After a series of Jackiw-Rebbi procedures, we end up with a chiral Majorana hinge modes on which the symmetry actions have explicit forms. Note that we will break TRS in the process, so eventually the system is a class D TSC with inversion symmetry.\\
The bulk Hamiltonian is:
\begin{equation}
    \hat{H}=\frac{1}{2}\sum_{\vec{k}}\hat{\psi}^\dag_{\vec{k}} H(\vec{k})\hat{\psi}_{\vec{k}},
\end{equation}
where
\begin{align}
    H(\vec{k})&=(-k_x\tau_x\sigma_z-k_y\tau_y+k_z\tau_x\sigma_x)+\lambda \tau_z,\nonumber \\
    \hat{\psi}^\dag_{\vec{k}}&=(c^{\dag}_{\vec{k},\uparrow},c^{\dag}_{\vec{k},\downarrow},c_{-\vec{k},\uparrow},c_{-\vec{k},\downarrow}).
\end{align}
The symmetries are represented as follows:
\begin{equation}
    \mathcal{T}=i\sigma_y \mathcal{K},\quad \mathcal{P}=\tau_x \mathcal{K},\quad \mathcal{I}=\tau_z.
\end{equation}
Note that here $\vec{\sigma},\vec{\tau}$ are Pauli matrices in different spaces. $\sigma$ is the spin space, and $\tau$ is the Nambu space. The 3D system lives inside a 3D ball, of which the boundary is $S^2$. Specifically we will pay attention to the gapless modes near $x^\pm=(\pm1,0,0)$, as these two points are related to each other by inversion. The strategy is as follows: first we perform Jackiw-Rebbi from 3D to 2D so that we end up with a system which is a 2D stacked system $H_{x^-}\oplus H_{x^+}$ with a 3D inversion symmetry which relates system $x^-$ and $x^+$ to each other; second we write a mass term $m(z)\Gamma$ in the stacked 2D system, and observe the behavior of the term under inversion symmetry. If $m(z)=-m(-z)$, then the stacked system hosts gapless mode along the hinge.\par 
The mass coefficient $\lambda\equiv \lambda(x)$ has the behaviour such that $\lambda=1$ inside the superconductor, and $\lambda=-1$ outside the superconductor. Let us investigate the surface modes near $x^+$. If we denote the surface eigenstate of the first quantized Hamiltonian as $\state{\varphi}$, then
\begin{equation}
    (i\tau_x\sigma_z\partial_x+\lambda(x)\tau_z)\state{\varphi}=0, \label{surface-eq:chiral}
\end{equation}
as we require the state to have no dispersion along the $x$-direction. The above equation is equivalent to
\begin{equation}
    \partial_x \state{\varphi}=i\lambda(x)\tau_x\tau_z\sigma_z\state{\varphi}=\lambda(x)\tau_y\sigma_z\state{\varphi}. \label{jr-eq:chiral}
\end{equation}
The above equation implies that $\state{\varphi}$ is an eigenstate of $\tau_y\sigma_z$. There are four eigenstates of $\tau_y\sigma_z$:
\begin{align}
    \state{+,1}&=(e^{-i\frac{\pi}{4}},0,e^{+i\frac{\pi}{4}},0)^\mathsf{T};\nonumber \\
    \state{+,2}&=(0,e^{+i\frac{\pi}{4}},0,e^{-i\frac{\pi}{4}})^\mathsf{T};\nonumber \\
    \state{-,1}&=(e^{+i\frac{\pi}{4}},0,e^{-i\frac{\pi}{4}},0)^\mathsf{T};\nonumber \\
    \state{-,2}&=(0,e^{-i\frac{\pi}{4}},0,e^{+i\frac{\pi}{4}})^\mathsf{T},
\end{align}
where the phases are added so that boundary excitations are explicitly Majorana fermions.\\
Using these eigenstates, the Eq.~\eqref{jr-eq:chiral} can be reduced to:
\begin{equation}
    \partial_x \state{\varphi,\pm}=\pm \lambda(x)\state{\varphi,\pm},
\end{equation}
for which the solutions are:
\begin{equation}
    \state{\varphi,\pm}=\exp [\pm\int^{x}_{x_0}dx'\lambda(x')]\state{\pm},
\end{equation}
where $x_0$ is a constant to fix the normalisation condition. We therefore see that the states with positive eigenvalues are the normalisable states near $x^+$.\\
We further define the following matrix:
\begin{equation}
    U=\frac{1}{\sqrt{2}}[\state{+,1},\state{+,2},\state{-,1},\state{-,2}].
\end{equation}
The first quantized Hamiltonian of the combined system is:
\begin{equation}
    H_{\text{comb}}(\vec{k})=U^\dag H(\vec{k})U=-k_ys_z\tilde{\sigma}_z+k_z\tilde{\sigma}_x,
\end{equation}
where ``comb'' stands for ``$x^+$ and $x^-$ combined". Note that we have omitted the exponential factor in the definition of $U$ for convenience. The exponential factor is only useful in telling us that the effective Hamiltonian is describing the physics near $x^+$, and writing it down explicitly helps us remove the $-k_x\tau_x\sigma_z+\lambda(x)\tau_z$ term after the projection. The $s$-space is now the space of $x^+$ and $x^-$.\par
We can now examine the representation of symmetries in this basis:
\begin{align}
    \mathcal{T}_{\text{comb}}&=U^\dag \mathcal{T} U=i\tilde{\sigma}_y\mathcal{K},\nonumber \\
    \mathcal{P}_{\text{comb}}&=U^\dag \mathcal{P} U=\mathcal{K},\nonumber \\
    \mathcal{I}_{\text{comb}}&=U^\dag \mathcal{I} U=-s_y\tilde{\sigma}_z. \label{symm_2D_chiral}
\end{align}
We conclude here, that we have obtained a stacked 2D system, of which the Hamiltonian is:
\begin{equation}
    \hat{H}_{\text{comb}}=\frac{1}{2}\sum_{\vec{k}}[\hat{\psi}^{x^+\dag}_{\vec{k}},\hat{\psi}^{x^-\dag}_{\vec{k}}] H_{\text{comb}}(\vec{k})\begin{bmatrix}
    \hat{\psi}^{x^+}_{\vec{k}}\\
    \hat{\psi}^{x^-}_{\vec{k}}
    \end{bmatrix},
\end{equation}
with symmetries having representations as in Eq.~\eqref{symm_2D_chiral}. We can have more explicit form of the spinor:
\begin{equation}
    \begin{bmatrix}
    \hat{\psi}^{x^+}_{\vec{k}}\\
    \hat{\psi}^{x^-}_{\vec{k}}
    \end{bmatrix}=\begin{bmatrix}
     \gamma^{x^+}_{-\vec{k},\uparrow}\\
    \gamma^{x^+}_{\vec{k},\downarrow}\\
    \gamma^{x^-}_{\vec{k},\uparrow}\\
    \gamma^{x^-}_{-\vec{k},\downarrow}
    \end{bmatrix}=U^\dag\hat{\psi}_{\vec{k}}.
\end{equation}
In terms of $c,c^\dag$, the Majoranas are:
\begin{align}
    \gamma^{x^+}_{-\vec{k},\uparrow}&=\frac{1}{\sqrt{2}}(e^{i\frac{\pi}{4}}c_{\vec{k},\uparrow}+e^{-i\frac{\pi}{4}}c^{\dag}_{-\vec{k},\uparrow})\nonumber \\
    \gamma^{x^+}_{\vec{k},\downarrow}&=\frac{1}{\sqrt{2}}(e^{-i\frac{\pi}{4}}c_{\vec{k},\downarrow}+e^{i\frac{\pi}{4}}c^{\dag}_{-\vec{k},\downarrow})\nonumber \\
    \gamma^{x^-}_{\vec{k},\uparrow}&=\frac{1}{\sqrt{2}}(e^{-i\frac{\pi}{4}}c_{\vec{k},\uparrow}+e^{i\frac{\pi}{4}}c^{\dag}_{-\vec{k},\uparrow})\nonumber \\
    \gamma^{x^-}_{-\vec{k},\downarrow}&=\frac{1}{\sqrt{2}}(e^{i\frac{\pi}{4}}c_{\vec{k},\downarrow}+e^{-i\frac{\pi}{4}}c^{\dag}_{-\vec{k},\downarrow}).
\end{align}
Let us add a surface perturbation which breaks TRS, i.e., a mass term on the whole surface of the 3D system that depends only on $z$. On our combined system, this perturbation is represented as $m(z)\Gamma$, where $\Gamma$ is some matrix. Such a term is $m(z)\tilde{\sigma}_y$. Also this term breaks TRS explicitly. By imposing $\mathcal{I}_{\text{comb}}m(z)\tilde{\sigma}_y \mathcal{I}^{-1}_{\text{comb}}=m(-z)\tilde{\sigma}_y$, we end up with $m(z)=-m(-z)$. Thus we can conclude that the class D TSC with inversion symmetry hosts hinge modes, i.e., it hosts second order topology, which is consistent with the previous work.\par
We now proceed to perform Jackiw-Rebbi procedures on the combined 2D system, so that symmetries on the 1D hinge modes will manifest.\par
The first quantized Hamiltonian with perturbation is:
\begin{equation}
    H_{\text{comb}}+\delta H_{\text{comb}}=-k_ys_z\tilde{\sigma}_z+k_z\tilde{\sigma}_x+m(z)\tilde{\sigma}_y,
\end{equation}
in which we assign the behavior of $m(z)$ to be
\begin{equation}
    m(z)=\begin{cases}
-1, \text{ if }z<0\\
0, \text{ if }z=0\\
+1, \text{ if }z>0
\end{cases}.
\end{equation}
If we denote the hinge eigenstate of the first quantized Hamiltonian as $\state{\varphi}$, then
\begin{equation}
    (-i\sigma_x\partial_z+m(z)\tilde{\sigma}_y)\state{\varphi}=0, \label{hinge-eq:chiral}
\end{equation}
as we require the state to have no dispersion along the $z$-direction. The above equation is equivalent to
\begin{equation}
    \partial_z \state{\varphi}=-im(z)\tilde{\sigma}_x\tilde{\sigma}_y\state{\varphi}=m(z)\tilde{\sigma}_z\state{\varphi}. \label{jr-eq:chiral_2}
\end{equation}
The above equation implies that $\state{\varphi}$ is an eigenstate of $s_0\tilde{\sigma}_z$. There are four eigenstates of $s_0\tilde{\sigma}_z$:
\begin{align}
    \state{+,1}&=(1,0,0,0)^\mathsf{T};\nonumber \\
    \state{+,2}&=(0,0,1,0)^\mathsf{T};\nonumber \\
    \state{-,1}&=(0,1,0,0)^\mathsf{T};\nonumber \\
    \state{-,2}&=(0,0,0,1)^\mathsf{T}.
\end{align}
Using these eigenstates, the Eq.~\eqref{jr-eq:chiral_2} can be reduced to:
\begin{equation}
    \partial_z \state{\varphi,\pm}=\pm m(z)\state{\varphi,\pm},
\end{equation}
for which the solutions are:
\begin{equation}
    \state{\varphi,\pm}=\exp [\pm\int^{z}_{z_0}dz'm(z')]\state{\pm},
\end{equation}
where $z_0$ is a constant to fix the normalisation condition. We therefore see that the states with positive eigenvalues are the normalisable states.\par
We further define the following matrix:
\begin{equation}
    U'=[\state{+,1},\state{+,2},\state{-,1},\state{-,2}].
\end{equation}
Note that $U'$ will take us to the space $\zeta\otimes \tilde{s}$ which can be read off by studying the basis. $\zeta$ is the space of normalisable states and non-normalisable states. We intend to keep the normalisable states as hinge states, therefore keep the $--$ block of the $\zeta$ space.\par
The first quantized Hamiltonian of hinge state is:
\begin{equation}
    {U'}^{\dag}(H_{\text{comb}}+\delta H_{\text{comb}})U'|_{--}={U'}^{\dag} (-k_ys_z\tilde{\sigma}_z) U'|_{--}=k_y\tilde{s}_z.
\end{equation}
The symmetries are:
\begin{align}
    \mathcal{P}_{\text{hinge}}&={U'}^{\dag} \mathcal{P}_{\text{comb}} U'|_{--}=\mathcal{K},\nonumber \\
    \mathcal{I}_{\text{hinge}}&={U'}^{\dag} \mathcal{I}_{\text{comb}} U'|_{--}=\tilde{s}_y. \label{symm_1D_chiralhinge}
\end{align}
And the spinor is:
\begin{equation}
    \hat{\Chi}_{\vec{k}}=(\chi^{x^+}_{\vec{k}},\chi^{x^-}_{-\vec{k}})^\mathsf{T}.
\end{equation}
The explicit form can be obtained in the following way:
\begin{equation}
    \hat{\Chi}_{\vec{k}}={U'}^{\dag}\begin{bmatrix}
    \hat{\psi}^{x^+}_{\vec{k}}\\
    \hat{\psi}^{x^-}_{\vec{k}}
    \end{bmatrix}|_{\text{lower}},
\end{equation}
since the normalisable states correspond to the $--$ block of the Hamiltonian. We can have more explicit forms from this expression:
\begin{align}
    \chi^{x^-}_{\vec{k}}&=\gamma^{x^-}_{-\vec{k},\downarrow}\nonumber \\
    \chi^{x^+}_{-\vec{k}}&=\gamma^{x^+}_{\vec{k},\downarrow}.
\end{align}
Therefore we can conclude that the 1D hinge Hamiltonian is:
\begin{equation}
    \hat{H}_{\text{hinge}}=\frac{1}{2}\sum_{\vec{k}}\hat{\Chi}^\dag_{\vec{k}} (k_y \tilde{s}_z)\hat{\Chi}_{\vec{k}}=\frac{1}{2}\sum_{k}\hat{\Chi}^\dag_{\vec{k}} H_{\text{hinge}}(\vec{k})\hat{\Chi}_{\vec{k}},
\end{equation}
with symmetries defined as in Eq.~\eqref{symm_1D_chiralhinge}.\par
We use $\theta$ to parametrise the full hinge which has periodicity $\pi$, and $x^-=0, x^+=\pi$. The action of inversion is therefore:
\begin{equation}
    I_{\text{hinge}}: \begin{bmatrix}
    \chi(\theta)\\
    \chi(\theta+\pi)
    \end{bmatrix}\mapsto \tilde{s}_y\begin{bmatrix}
    \chi(\theta)\\
    \chi(\theta+\pi)
    \end{bmatrix}=\begin{bmatrix}
    -i\chi(\theta+\pi)\\
    i\chi(\theta)
    \end{bmatrix}.
\end{equation}
We conclude that, the 3D class D TSC with inversion symmetry hosts a single chiral Majorana hinge mode $\chi$ on the surface with symmetries defined as above. 
 
 \section{Third-order class DIII inversion-symmetric superconductor }
 In this appendix, we show that the surface zero modes of 3D class DIII topological superconductor with inversion are stable to weakly-interacting surface perturbations.\par
We proceed to describe the weakly-interacting surface perturbation. Specifically, if we specify that the great circle connecting the antipodal corner modes as the equator, then we paste one copy of 2D TRS invariant topological superconductor on the northern hemisphere surface and another copy of 2D TRS invariant topological superconductor on the southern hemisphere surface, and these two copies are related to each other by inversion symmetry. To show that the original point modes are stable under this perturbation, it is sufficient to show that the pasted 2D model will not host any new corner modes on the equator, as these new corner modes will be able to be used to 'annihilate' the original point modes.\par
Our effectively 2D model, which can be called N-S model, has the following Hamiltonian:
\begin{equation}
H=(-k_x\sigma_z \tau_x-k_y\tau_y+\lambda \tau_z)\zeta_0,
\end{equation}
where $\zeta$ denotes the N-S orbital space. The symmetries in this model are as follows:
\begin{equation}
 \mathcal{T}=\sigma_y \mathcal{K},\quad  \mathcal{P}=\tau_x \mathcal{K},\quad \mathcal{I}=\zeta_x\tau_z,
\end{equation}
note that inversion switches the N-S orbitals. \par
The 1D edge of the N-S model, i.e. the equator, can be completely gapped out without leaving point modes behind. This can be shown by introducing the following gapping term in the bulk:
\begin{equation}
\delta H=m_{\vec{r}} \sigma_x \tau_x \zeta_y,
\end{equation}
which is TRS invariant and PHS respecting. At the same time, by requiring $\mathcal{I}\delta H(\vec{r})\mathcal{I}^{-1}=\delta H(-\vec{r})$, we necessarily arrive at $m_{\vec{r}}=m_{-\vec{r}}$. To see what this means on the equator, we invoke the projection procedures as before. Upon projection, the 1D Hamiltonian and the gapping term become:
\begin{equation}
h=-k_S\sigma_z\zeta_0, \delta h=-m_{\vec{r}}(\hat{n}_{\vec{r}}\times \vec{\sigma})_z\zeta_y,
\end{equation}
with symmetries:
\begin{equation}
 \mathcal{T}_S=\sigma_y \mathcal{K},\quad \mathcal{P}_S=-(\hat{n}_{\vec{r}}\cdot\vec{\sigma})\sigma_y \mathcal{K},\quad \mathcal{I}_S=-\zeta_x.
\end{equation}
Note that here $k_S$ denotes momentum perpendicular to $\hat{n}_{\vec{r}}$. Here we can see again, by requiring $\mathcal{I}_S\delta h(\vec{r})\mathcal{I}^{-1}_S=\delta h(-\vec{r})$, we necessarily arrive at $m_{\vec{r}}=m_{-\vec{r}}$.\par
Thus we have proved that the edge of N-S model can be completely gapped without breaking symmetries or leaving corner modes, and consequently, the corner modes that arise from the third order topology of the 3D class DIII topological superconductor with inversion symmetry are stable under such surface perturbation.

\bibliography{bibliothek}
\end{document}